\documentclass[a4paper]{llncs} 
\usepackage{amssymb}
\usepackage{graphicx}
\usepackage{url}
\usepackage{multirow}
\usepackage{verbatim}
\usepackage[english]{babel}
\usepackage{rotating}
\usepackage{authblk}
\usepackage{alltt}
\usepackage{moreverb}
\usepackage{theorem}
\usepackage{amsmath}
\usepackage{multirow}
\usepackage[table]{xcolor}
\usepackage{array,ragged2e}
\usepackage{longtable}
\usepackage{pdflscape}

\usepackage{lipsum}
\usepackage{booktabs}
\usepackage{dblfloatfix}
\usepackage{multirow}
\usepackage{verbatim}
\usepackage{alltt}
\usepackage{afterpage}

\usepackage{caption}

\usepackage{pgf-pie}

\usepackage{tikz}

\usepackage{hyperref}

\usepackage{longtable}
 
\usepackage{pgfplots, pgfplotstable}

\usepackage{xcolor}
\usepackage{pgfplots}

\definecolor{bblue}{HTML}{4F81BD}
\definecolor{rred}{HTML}{C0504D}
\definecolor{ggreen}{HTML}{9BBB59}
\definecolor{ppurple}{HTML}{9F4C7C}

\usepackage{xcolor}
\newlength\MAX  
\newcommand*\Chart[1]{#1~\rlap{\textcolor{black!20}{\rule{\MAX}{2ex}}}\rule{#1\MAX}{2ex}}

\usepackage{amsmath,amssymb}
\usepackage{array}
\usepackage{xcolor}

\usetikzlibrary{positioning,shadows}

\newif\ifpienumberinlegend
\pgfkeys{/number in legend/.code=
    \expandafter\let\expandafter\ifpienumberinlegend
    \csname if#1\endcsname
    \ifpienumberinlegend

    \def\beforenumber##1\afternumber{}%
    \fi,
    /number in legend/.default=true
}
\begin{document}

\mainmatter  

\title{A Core Ontology for Privacy Requirements Engineering}

\titlerunning{A Core Ontology for Privacy Requirements Engineering}

%
%

\author{Mohamad Gharib\inst{1}  \and    John Mylopoulos\inst{2}}
\authorrunning{Gharib \& Mylopoulos}



\institute{University of Florence - DiMaI, Viale Morgagni 65, Florence, Italy \\   \email{mohamad.gharib@unifi.it}
\and     University of Trento - DISI, 38123, Povo, Trento, Italy\\    	\email{john.mylopoulos@unitn.it}}

%
%

\toctitle{Lecture Notes in Computer Science}
\tocauthor{Authors' Instructions}
\maketitle

\begin{abstract}

Nowadays, most companies need to collect, store, and manage personal information in order to deliver their services. Accordingly, privacy has emerged as a key concern for these companies since they need to comply with privacy laws and regulations. To deal with them properly, such privacy concerns should be considered since the early phases of system design. Ontologies have proven to be a key factor for elaborating high-quality requirements models. However, most existing work deals with privacy as a special case of security requirements, thereby missing essential traits of this family of requirements.  In this paper, we introduce COPri, a Core Ontology for Privacy requirements engineering that adopts and extends our previous work on privacy requirements engineering ontology that has been mined through a systematic literature review. Additionally, we implement, validate and then evaluate our ontology.

\section*{Keywords} Privacy Ontology, Privacy Requirements, Privacy by Design (PbD), Requirements Engineering, Socio-technical Systems

\end{abstract}

\section{Introduction}

Nowadays, most companies collect, store, and manage personal information (e.g., information about customers, citizens, etc.) to deliver their services. These companies need to protect the privacy of personal information not only for maintaining their credibility and repetition but also to comply with various privacy laws and regulations \cite{gharib2016privacy}.  More specifically, most developed countries have developed various laws and regulations to govern the use of personal information. For instance, the General Data Protection Regulation (GDPR) \cite{GDBR2016regulation} has been recently developed by the European Union (EU) with the aim to safeguard the use of personal information among all EU member states. Moreover, the Australia Government issued the Privacy Act 1988 \cite{privacylawAus1988}, which include a set of privacy rights known as the Information Privacy Principles (IPPs). Canada has developed the Personal Information Protection and Electronic Documents Act (PIPEDA) \cite{privacylawcanada2000} that regulates how personal information can be collected, used and disclosed. In the United States, more domain-specific laws have been developed (e.g., HIPAA \cite{hippa96} for healthcare domain, the Financial Services Modernization Act \cite{financeprivacylaw2002}, etc.) 

Failing to comply with privacy laws and regulations results in huge monetary sanctions, which companies want to avoid \cite{Gharib2017ER}. Accordingly, privacy has become a main concern for system designers. In other words, dealing with privacy-related concerns is a must these days because privacy breaches may have severe consequences \cite{gellman2002privacy,hong2004privacy}.  In particular, the absence of appropriate privacy protection mechanisms may lead to privacy breaches, which impose huge direct cost \cite{acquisti2006there,gellman2002privacy}, as well as long-term consequences \cite{campbell2003economic,cavusoglu2004effect} such as having one's personal information in the wrong hands \cite{Margulis2003}. However, most of these breaches can be avoided if the privacy requirements of the system-to-be were captured properly during system design (e.g., Privacy by Design (PbD)) \cite{kalloniatis2008addressing,labda2014modeling}.  Nevertheless, most existing work on privacy requirements often deal with them either as non-functional requirements (NFRs) with no specific techniques on how such requirements can be met \cite{mouratidis2007secure}, or as security requirements (e.g., \cite{zannone2006requirements,kalloniatis2008addressing}), i.e.,  focusing mainly on confidentiality and overlooking important privacy aspects such as anonymity, pseudonymity, unlinkability, unobservability, etc.

On the other hand, privacy is one of the few concepts that has been studied across many discipline including law \cite{warren1890right}, sociology \cite{westin1968privacy,Etzioni2004}, psychology \cite{altman1976conceptual}, and information systems \cite{culnan1999information}.  Although it has been studied for more than a century, it is still elusive and vague concept to grasp \cite{solove2002conceptualizing,solove2006taxonomy,kalloniatis2008addressing}.  Despite this, numerous attempts have been made by scholars to clarify the concept by linking it to more refined concepts such as secrecy, confidentiality, anonymity, pseudonymity, unlinkability, unobservability, control of personal information \cite{solove2006taxonomy,zwick2004whose,Pfitzmann2010}, or to solitude, intimacy, anonymity, and reserve as in \cite{westin1968privacy}.  Other studies suggest that the notion of risk is also related to privacy as the loss of information control implies risk \cite{phelps2000privacy,sheehan2000dimensions,krasnova2010online}.  While Awad and Krishnan \cite{Awad2006} investigated how transparency can influence privacy. However, there is no consensus on the definition of these concepts or which of them should be used to analyze privacy \cite{solove2006taxonomy}.  

In addition, many of these concepts are overlapping, which contributes to the confusion while dealing with privacy \cite{Dinev2013}. This has resulted in much confusion among designers and stakeholders, and has led in turn to wrong design decisions. Ontologies have proven to be a key success factor for eliciting high-quality requirements, as they reduce the conceptual vagueness and terminological confusion by providing a shared understanding of the related concepts between the designers and stakeholders of the system \cite{uschold1996ontologies,kaiya2006using,dzung2009ontology,souag2015security}. In this context, a well-defined ontology that captures key privacy-related concepts and relationships could solve this problem.

Privacy is a social concept \cite{Margulis2003,Gharib2017ER}. Accordingly, the privacy ontology should conceptualize privacy requirements in their social and organizational context \cite{Gharib2017ER}. In other words, the ontology should consider not only the technical aspects of privacy but also its related social and organizational aspects. Since most systems these days are socio-technical systems consisting not only of technical components but also of humans along with their interrelationships, where different kinds of vulnerabilities might manifest themselves \cite{liu2003security,gharib2016privacy}.  More specifically, focusing on the technical aspects and leaving the social and organizational aspects outside the system's boundary leaves the system open to different kinds of vulnerabilities that might manifest themselves in the social interactions and/or the organizational structure of the system. 

In previous research \cite{Gharib2017ER}, we worked toward addressing this problem by proposing an ontology for privacy requirements that has been mined through a systematic literature review. In this paper, we extend the ontology proposed in \cite{Gharib2017ER} with new and more refined concepts concerning both personal information and privacy requirements. Moreover, we implement the ontology, apply it to an Ambient-Assisted Living (AAL) illustrating example, and then  validate it by querying the ontology instance (e.g., the AAL example) depending on a set of competency questions. Finally, we evaluate the ontology against common pitfalls in ontologies with the help of some tools,  lexical semantics experts, and privacy and security researchers.

The rest of the paper is organized as follows; Section (\textsection 2) presents the AAL example that is used to illustrate our work, and we describe the process we followed for developing the COPri ontology in Section (\textsection 3).  Section (\textsection 4) describes the conceptual model of COPri, and we implement and validate the ontology in Section (\textsection 5) and (\textsection 6) respectively. We evaluate the ontology in Section (\textsection 7), and we discuss threats to its validity in Section (\textsection 8). Related work is presented in Section (\textsection 9), and we conclude and discuss the future work in Section (\textsection 10).

\section{Illustrating example: the Ambient-Assisted Living (AAL) System}

Longevity among the elderly has result in many challenges for society and the health care system as well, such as increasing in age-related diseases (e.g., Alzheimer, diabetes, etc.), which in turn leads to a shortage of caregivers \cite{rashidi2013survey}.  But this is not the only problem since most older people (around 89\%) prefer to stay at their own homes \cite{ziefle2011medical,rashidi2013survey},  and given the costs of home care nursing, it is imperative to develop technologies that help older people to age in place \cite{ziefle2011medical}. 

AAL systems sound to be an appropriate solution to this problem. AAL systems rely on monitoring and actuating devices to shift some of the healthcare services from a hospital-centered to a patient-centric treatment \cite{brandao2012abstracting}. In other words,  instead of being measured face-to-face, a patient's health status can be sensed remotely, continuously, and in real time, and then such information is processed and transferred to a hospital or health care center \cite{he2012distributed}. Moreover, AAL technologies facilitate communication among physicians and patients, and allows for discussing medical data and negotiating treatment procedure remotely \cite{beul2011s}. This decrease both the costs of health care services and also the workload of medical practitioners \cite{yusof2002role,miller2003technical,ziefle2011medical}. However, numerous studies showed that privacy is one of the most highlighted criticisms for such technology \cite{Hong2004}.

\begin{figure*}[!t]
\centering
\includegraphics[width= 0.9 \linewidth]{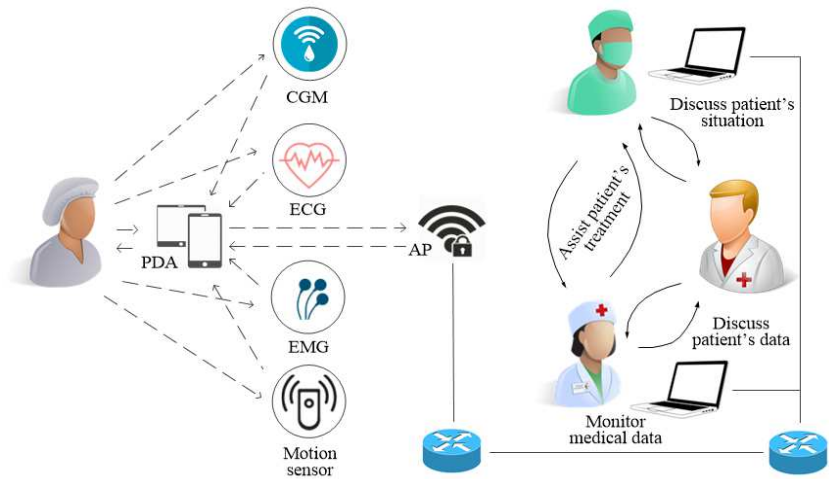}
\caption{Simplified representation of the AAL system }
\label{fig:aal}
\end{figure*}

Our motivating example concerns an old person called Jack that suffers from diabetes disease. Jack lives in a home that is equipped with AAL system, which provides an appropriate environment for Jack to live normally. In particular, the AAL system depends on various interconnected body sensors (e.g.,  electroencephalography (ECG), electromyography (EMG), Continuous Glucose Monitoring (CGM), location, and motion sensors) that collect various information concerning Jack's vital signs, location, and activities. This information is transmitted to Jack's Personal Digital Assistant (PDA) that assesses his health situation and provide required notification accordingly. 

Jack' PDA also forward the information to a nearby caring center, where a virtual nurse called Sarah can monitor such information, and she can also monitor some of Jack's activities (e.g., watching tv, sleeping, preparing or having a meal, etc.) by collecting location and motion activities. Sarah can detect unusual situations and react accordingly, she also has access to all Jack's health records and she may contact the required medical professional (e.g., General Practitioner (GP), consulting physicians) that might be needed depending on Jack's situation. Jack, like many other users, wants to preserve his privacy by controlling what is collected and shared of his personal information, who is using such information, and for which reasons.
 
Figure \ref{fig:aal}  shows a simplified representation of the AAL system that Jack depends on, which adopts a three-tier Body Area Network (BAN) communication architecture \cite{Networks2010}. Such architecture classifies the BAN communications into three types: \textit{1- intra-BAN} that has a range of about two meters around the human body and covers communications between body sensors (e.g., ECG, EMG, CGM, and motion sensor) and PDA; \textit{2- inter-BAN} covers communications between a PDA and one or more access points (APs); and \textit{3- beyond-BAN} connects the APs to the internet and other networks. This architecture helps in better understanding and dealing with privacy requirements/concerns.

\section{The process for developing the COPri ontology}

The process for developing COPri has been developed based on \cite{gomez1996towards,uschold1996building,fernandez1997methontology} following the five principles proposed by Gruber  \cite{gruber1995toward} (e.g., clarity, coherence, extendibility,  minimal encoding bias, and minimal ontological commitment). The process is depicted in Figure \ref{fig:process}, and it is composed of five main steps:

\begin{itemize}

\item \textit{Step 1. scope \& objective identification} aims at identifying the scope of the ontology, the purposes it will be used for, and its intended users \cite{uschold1996building,fernandez1997methontology}. As previously highlighted, there is a need for addressing privacy concerns during the system design (e.g., Privacy by Design (PbD) \cite{kalloniatis2008addressing,labda2014modeling}). Nevertheless, based on the results of our systematic literature review \cite{Gharib2017ER}, most existing studies miss several key privacy concepts and relationships. Therefore, it is almost impossible to address main privacy concerns during the system design. To this end, COPri aims at assisting software engineers while designing privacy-aware systems that belong to various domains by providing a generic and expressive set of key privacy concepts and relationships, which enable for capturing privacy requirements of the system-to-be in their social and organizational context.

\item \textit{Step 2. Knowledge acquisition}  aims at identifying and collecting knowledge needed for the construction of the ontology.  We have conducted a systematic literature review with a main purpose of identifying the key concepts and relationships for capturing privacy requirements\footnote{A detailed version of the systematic literature review can be found at \cite{gharib16arXiv}}. In particular, five electronic database sources have been used for the acquisition of knowledge.  240 relevant papers have been returned, among which we have selected 34 after removing duplicated papers and applying several selections and quality assessment criteria. Then, we have analyzed the contents of selected studies identifying 38 privacy related concepts and relationships\footnote{In the case of multiple synonyms, some were omitted}, which have been grouped into four main groups based on their type:  17 \textit{organizational concepts} capture social and technical aspects of the system-to-be; 9 concepts to capture \textit{risks} that might endanger privacy requirements; 5 \textit{treatment concepts} capture countermeasure techniques to mitigate risks to privacy needs; and 7 \textit{privacy concepts} capture the stakeholders privacy requirements/needs.

\begin{figure*}[!t]
\centering
\includegraphics[width= 0.9 \linewidth]{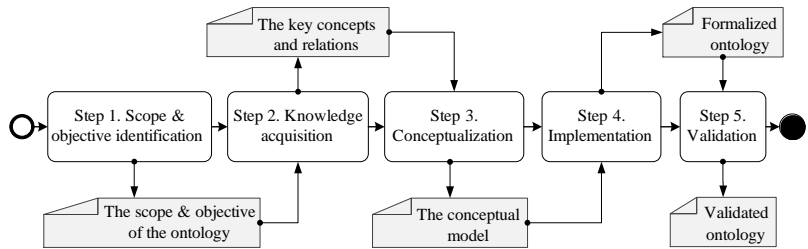}
\caption{The process for developing the COPri ontology}
\label{fig:process}
\end{figure*}

\item  \textit{Step 3. Conceptualization} aims at structuring the acquired knowledge into a conceptual model that captures the key concepts of the ontology along with their interrelationships \cite{fernandez1997methontology}. In \cite{Gharib2017ER}, we have built an ontology based on the 38 selected key concepts and relationships. In his paper, we extend this ontology with more refined concepts concerning both personal information and privacy requirements.  Additionally, we conducted a survey to collect feedback from privacy and security researchers to evaluate the completeness of our proposed ontology, i.e., determine whether the selected concepts and relationships are capable of properly dealing with privacy requirements or they need to be extended or refined. The feedback confirmed that most of the concepts and relationships are appropriate and the ontology is capable of capturing privacy requirements. Some feedback suggested to refine, include or exclude some concepts/relationships, we took these suggestions into account while developing the final ontology. A detailed description of the resulting ontology (COPri) is provided in the next section.

\item \textit{Step 4. Implementation} aims at codifying the ontology in a formal language. This requires an environment that guarantees the absence of lexical and syntactic errors from the ontology; translators, to guarantee the portability of the definitions into other target languages; and an automated reasoner to detect incompleteness, inconsistencies and redundant knowledge \cite{uschold1995towards}.  Although there exist several environments for developing (codifying) ontologies (NeOn Toolkit \cite{Haase2008}, OntoEdit  \cite{Sure2002a}, SWOOP  \cite{Kalyanpur2006}, prot\'eg\'e \cite{Prot2009}).  We have chosen Prot\'eg\'e\footnote{\url{http://protege.stanford.edu/}} that is a set of open-source and domain-independent ontology design software developed in Stanford Medical Informatics. Prot\'eg\'e can be used easily for creating, modifying, visualizing and checking the consistency of ontology. Moreover, the reasoner can be used to automatically compute a classification hierarchy (\textit{inferred hierarchy}) based on a manually constructed class hierarchy that is called the \textit{asserted hierarchy}. In addition, prot\'eg\'e offers several useful plug-ins for visualizing ontology, and most importantly it offers a plug-in for using SPARQL (Protocol and RDF Query Language) to extract knowledge from ontology through defined queries and rules \cite{prud2006sparql}. The implementation of COPri is discussed in section 5.

\item \textit{Step 5. Validation,} aims at ensuring that the resulting ontology meets the needs of its usage, i.e., the ontology corresponds to the system, which it is supposed to represent \cite{fernandez1997methontology}. According to \cite{uschold1996building}, informal and formal questions/queries can be used to validate ontology. Following \cite{Fox1993,Dong2007}, we validated COPri after applying it to the AAL illustrating example by querying the ontology instances depending on Competency Questions (CQs), and then verifying the correctness of the results of such queries.  More specifically, the CQs are used to evaluate whether the ontology captures enough detailed information about the targeted domain to fulfill the needs of its intended use. The validation of  COPri is discussed in more details in section 6.

\end{itemize}

\section{The conceptual model of COPri}

In this section, we present the conceptual model of COPri in terms of its concepts and relationships.  Figure \ref{fig:COPri} shows the meta-model of COPri as a UML class diagram. The concepts of COPri are organized into four main dimensions:

\begin{description}

\item[\textbf{Organizational dimension:}] proposes concepts to capture the social and technical components of the system in terms of their capabilities, objectives, and dependencies.

\item[\textbf{Risk dimension:}] proposes concepts to capture risks that might endanger privacy needs at the social and organizational levels.

\item[\textbf{Treatment dimension:}] proposes concepts to capture countermeasure techniques to mitigate risks to privacy needs.

\item[\textbf{Privacy dimension:}] proposes concepts to capture the stakeholders' (actors) privacy requirements/needs concerning their personal information.

\end{description}

\textbf{(1) Organizational dimension,} includes concepts for capturing the organizational aspects of the system, which are further organized into several categories such as agentive, intentional and informational entities, social dependencies and social trust. In what follows, we define each of these categories in terms of their concepts and relationships. 

\underline{\textbf{Agentive entities:}} captures  the active entities of the system, we have three concepts along with two relationships:

\begin{figure*}[!t]
\centering
\includegraphics[width=0.91 \textheight,angle=90]{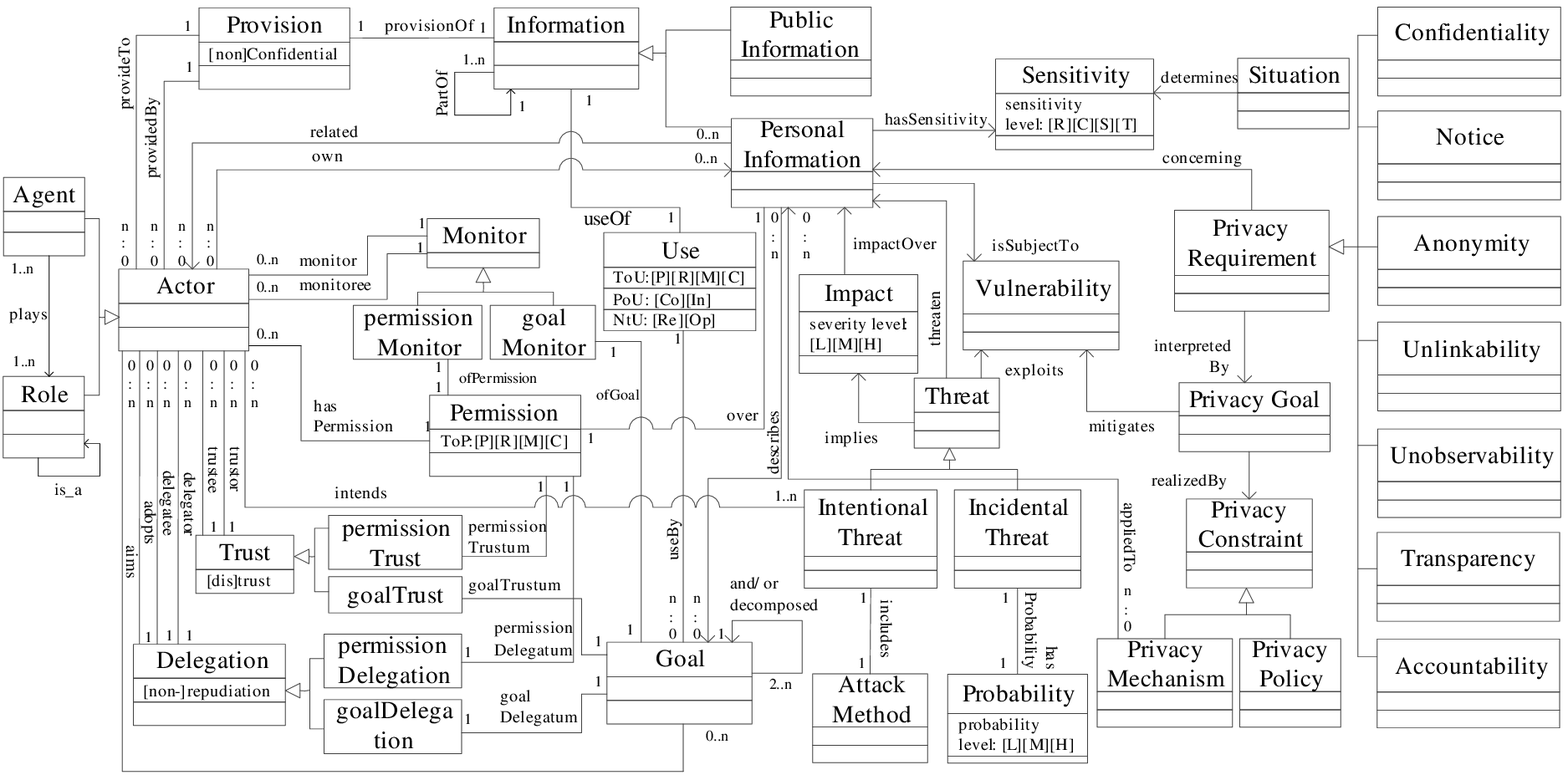}
\caption{The conceptual model of COPri}
\label{fig:COPri}
\end{figure*}

\begin{description}

\item[\textbf{Actor}] represents an autonomous entity that has intentionality and strategic goals within the system, and it covers two entities: a role and an agent:  

\begin{description}

\item[\textbf{Role}] represents an abstract characterization of an actor in terms of a set of behaviors and functionalities within some specialized context. A role can be a specialization (\textbf{is\_a} relationship) of one another.

\item[\textbf{Agent}] represents an autonomous entity that has a specific manifestation in the system. An agent can \textbf{plays} a role or more within the system, where an agent inherits the properties of the roles it plays.
\end{description}
\end{description}

\underline{\textbf{Intentional entities:}} captures the objectives that the actors aim to achieve. Therefore, we adopted the goal concept as well as and/or decomposition (refinement) relationships to represent such objectives. 

\begin{description}

\item[\textbf{A goal}] is a state of affairs that an actor aims to achieve. When a goal is too coarse to be achieved, it can be refined through \textit{and/or-decompositions} of a root goal into finer sub-goals.  

\item[\textbf{and-decomposition}] implies that the achievement of the root-goal requires the achievement of all of its sub-goals. 

\item[\textbf{or-decomposition}] is used to provide different alternatives to achieve the root goal, and it implies that the achievement of the root-goal requires the achievement of any of its sub-goals. 

\end{description}

\underline{\textbf{Informational entities:}} capture  the \textit{Information} related concepts and relationships:

\begin{description}

\item[\textbf{Information}] represents a statement provided or learned about something or someone. Information can be atomic or composite (composed of several parts), and we rely on \textit{partOf} relationship to capture the relationship between an information entity and its sub-parts. Moreover, we differentiate between two types of information:

\begin{description}

\item[\textbf{Public information,}] any information that cannot be \textit{related} (directly or indirectly) to an identified or identifiable legal entity.

\item[\textbf{Personal information,}] any information that can be \textit{related} (directly or indirectly) to an identified or identifiable legal entity (e.g., names, addresses, medical records) \cite{braghin2008introducing,van2003handbook}. 

\end{description}
\end{description}

Several researchers have advocated that not all personal information has the same sensitivity levels (e.g., \cite{van2003handbook,dritsas2006knowledge,labda2014modeling}). Moreover, various sensitivity levels and categories for personal information have been proposed (e.g., \cite{comber1969management,ellis1972privacy,Turn1976,kosa2011measuring}). To this end, we include \textit{sensitivity level} concept that \textit{personal information} has in our ontology.  Based on \cite{Turn1976}, we adopt four different sensitivity levels ordered as  \textit{(R)estricted}, \textit{(C)onfidential}, \textit{(S)ensitive}, and \textit{Secre(T)}, where \textit{Secre(T)} is the most sensitive. Personal information with different sensitivity levels may have different privacy requirements, i.e., sensitivity levels can be used to facilitate the identification of privacy requirements.

On the other hand,  numerous works (e.g., \cite{Nissenbaum2004a,Barth2006,Omoronyia2013}) have linked the sensitivity of personal information to when and where such information has been collected and for what purposes, i.e., the context/state of affairs related to such information. Thus, we adopt the concept of \textit{situation} as a mean to determine the sensitivity level of personal information, where a \textit{situation} can be defined as a partial state of affairs in terms of things that exist in that state, their properties, and interrelations \cite{Horkoff2014}.

\underline{\textbf{Use}} is a relationship between a goal and information, and it has three attributes:

\begin{description}

\item[\textbf{Type of Use (ToU),}] our ontology provide four different types of use:

\begin{description}
\item[\textbf{Produce}] indicates that information is created by a goal;
\item[\textbf{Read}] indicates that information is consumed by a goal;
\item[\textbf{Modify}] indicates that information is modified/altered by a goal;
\item[\textbf{Collect}] indicates that information is acquired by a goal.
\end{description}

\item[\textbf{Need to Use (NtU)}] captures the necessary of use, and we differentiate between two main types:

\begin{description}
\item[\textbf{Require}] indicates that the use of information is required for the goal achievement  \cite{gharibRefsq2015};
\item[\textbf{Optional}] indicates that information is not required for the goal achievement \cite{gharibRefsq2015}.
\end{description}

\item[\textbf{Purpose of Use (PoU),}] we differentiate between two main categories of purposes of use:

\begin{description}
\item[\textbf{Compatible}] indicates that the purpose for which information is used is compliant with the rules that guarantee the best interest of its owner;
\item[\textbf{Incompatible}] indicates that the purpose for which information is used is not compliant with the rules that guarantee the best interest of its owner.
\end{description}
\end{description}

\underline{\textbf{Describes}} is a relationship where information characterizes a goal (activity) while it is being pursued by some actor.

\underline{\textbf{Information ownership \& Permissions:}} capture the relationships among personal information, the legal entities who own them, and how such entities control the use of such information by others.

\begin{description}

\item[\textbf{Own}] indicates that an actor is the legitimate owner of information, where information owner has full control over the use of information it owns.

\item[\textbf{Permission}] is consent that identifies a particular use of a particular information in a system \cite{sandhu1996role}. Information owner (data subject\footnote{ Information owner and data subject are synonyms in this paper}) controls the use of its own information depending on \textit{permissions over} such information. In COPri, a permission has a type that is (P)roduce, (R)ead, (M)odify and (C)ollect, which cover the four relationships between goals and information that our ontology proposes.

\end{description}

\underline{\textbf{Entities interactions:}} capture the interactions/dependencies among actors of the system concerning their objectives and entitlements. The ontology adopts three types of interactions:

\begin{description}

\item[\textbf{Information provision}] captures the transmission of information (\textit{provisionOf}) by an actor (\textit{provisionBy}) to  another one (\textit{provisionTo}), where the source of the provision relationship is the provider and the destination is the requester. Moreover, information provision has a type that can be either \textit{confidential} or \textit{nonConfidential}, where the former guarantee the confidentiality of the transmitted information, while the last does not.

\item[\textbf{Delegation}] indicates that actors can delegate obligations and entitlements to one another, where the source of delegation called the delegator, the destination is called delegatee, and the subject of delegation is called delegatum. The concept of \textit{delegation} is further specialized into two concepts: \textit{Goal delegation,} where the delegatum is a goal; and \textit{Permission delegation,} where the delegatum is a permission.

\item[\textbf{Adoption}] is considered as a key component of social commitment, and it indicates that an actor accepts to take responsibility for the delegated objectives and/ or entitlements from another actor \cite{castelfranchi1998modelling}.

\end{description}

\underline{\textbf{Entities social trust:}} the need for trust arises when actors depend on one another for goals or permissions since such dependencies might entail risk  \cite{chopra2003trust,gharib2015analyzing}.  Therefore, our ontology adopts the concept of \textit{trust} to capture the actors' expectations of one another concerning their delegations. The source of trust called the trustor, the destination is called trustee, and the subject of trust is called trustum.   \textit{Trust} has a type that can be either: 

\begin{description}
\item[\textbf{Trust}] means the trustor expect that the trustee will behave as expected considering the trustum (e.g., a trustee will achieve the delegated goal, or it will not misuse the delegated permission).

\item[\textbf{Distrust}] means the trustor expect that the trustee will not behave as expected considering the trustum (e.g., a trustee will not achieve the delegated goal, or it will misuse the delegated permission).
\end{description}

The concept of \textit{Trust} is further specialized into two concepts \textit{GoalTrust,} where the trustum is a goal; and \textit{PermissionTrust,} where the trustum is a permission.

\underline{\textbf{Monitoring:}} can be defined as the process of observing and analyzing the performance of an actor in order to detect any undesirable performance \cite{guessoum2004monitoring}. We adopt the concept of \textit{monitoring} to compensate the lack of trust or distrust in the trustee concerning the trustum \cite{gans2001modeling,zannone2006requirements}, where the source of monitoring is called the monitor, the destination is called monitoree. 

The concept of \textit{monitor} is further specialized into two concepts \textit{GoalMonitor,} where the subject of the monitoring is a goal; and \textit{PermissionMonitor,} where the subject of the monitoring is a permission.

\textbf{(2) Risk dimension,} includes risk related concepts along with their interrelationships (e.g., threat, vulnerabilities, attack, etc.) concerning personal information. In what follows, we define each of these concepts and their interrelationships: 

\begin{description}

\item[\textbf{A vulnerability}] is a weakness in the system that can be \textit{exploited} by a \textit{threat} \cite{rostad2006extended,mayer2009model,singhal2010ontologies}.

\item[\textbf{A threat}] is a potential incident that \textit{threaten} personal information by \textit{exploiting} a \textit{vulnerability} concerning such information \cite{mayer2009model,singhal2010ontologies,kang2013security}. A \textit{threat} can be either natural (e.g. disaster), accidental (e.g. hardware or software failure), or intentional (e.g. theft of personal information) \cite{velasco2009modelling,souag2015security}. COPri differentiates between two types of threat:

\item[\textbf{Incidental threat}] is  a casual, natural or accidental threat that is not caused by a \textit{threat actor} nor require an \textit{attack method}. \textit{Incidental threat} has a \textit{probability} that measures the likelihood of its occurs, and it is is characterized by three different values \textit{high}, \textit{medium} or \textit{low}.

\item[\textbf{Intentional threat}] is a threat that require a \textit{threat actor}  and \textit{includes} a presumed \textit{attack method} \cite{lin2003introducing,massacci2011extended}.

\item[\textbf{Threat actor}] is an actor that intends to achieve an \textit{intentional threat} \cite{rostad2006extended,mayer2009model,elahi2009modeling}.

\item[\textbf{Attack method}] is a standard means by which a \textit{threat actor} carries out an \textit{intentional threat} \cite{mayer2009model,elahi2010vulnerability,souag2015security}.

\item[\textbf{Impact}] is the consequence of the \textit{threat} over the personal information. An \textit{impact} has a \textit{Severity} that captures the level of the impact \cite{wang2009ovm,souag2015security}, which is characterized by    \textit{high}, \textit{medium} or \textit{low}.

\end{description}

\textbf{(3)  Treatment dimension,} includes countermeasure concepts to mitigate risks. COPri proposes a high abstraction level concepts to capture the required protection/treatment level (e.g., privacy goal), which can be refined into concrete protection/treatment constraints (e.g., mechanisms or policies) that can be implemented. The concepts of the treatment dimension are:

\begin{description}

\item[\textbf{A privacy goal}] defines an aim to counter threats and prevents harm to personal information by satisfying privacy criteria concerning such information. 

\item[\textbf{A privacy constraint}] is a design restriction that is used to realize/satisfy a privacy goal, constraints can be either a privacy policy or privacy mechanism.

\item[\textbf{A privacy policy}]  is a privacy statement that defines the permitted and/or forbidden actions to be carried out by actors of the system toward information.

\item[\textbf{A privacy mechanism}] is a concrete technique to be implemented for helping towards the satisfaction of privacy goal. Some mechanisms can be directly \textit{applied to} \textit{personal information} (e.g., anonymity, unlinkability).

\end{description}

\textbf{(4)  Privacy dimension,} introduce concepts to capture the actors' privacy requirements/needs concerning their personal information. The concepts of the privacy dimension are:

\begin{description}

\item[\textbf{Privacy requirement}]  that is used to capture information owners' privacy needs at a high abstraction level \textit{concerning} their \textit{personal information}. \textit{Privacy requirements} are \textit{interpretedBy} \textit{privacy goals}. Moreover, \textit{privacy requirement} is further specialized into seven more refined concepts:

\item[\textbf{Confidentiality}] means personal information should be kept secure from any potential leaks and improper access \cite{solove2006taxonomy,dritsas2006knowledge,labda2014modeling}. We rely on the following principles to analyze confidentiality: 

\begin{description}

\item[\textbf{Non-disclosure,}] personal information can only be disclosed if the owner's consent is provided, i.e., the disclosure of the personal information should be under the control of its legitimate owner \cite{solove2006taxonomy,dritsas2006knowledge,braghin2008introducing,labda2014modeling}. 

Note that \textit{non-disclosure} also covers information provision (e.g., confidential information provision). Therefore, \textit{non-disclosure} can be analyzed depending on the existence of read permission as well as the confidentiality of information provision. 

\item[\textbf{Need to Know (NtK),}] personal information can only be used if it is strictly necessary for completing a certain task \cite{labda2014modeling}.   \textit{NtK} can be analyzed depending on \textit{Need to Use  (NtU)} that captures the necessary of use,  i.e., if the type of \textit{NtU} is \textit{optional} (i.e., not required) a violation can be raised.

\item[\textbf{Purpose of Use (PoU),}] personal information can only be used for specific, explicit, legitimate purposes and not further used in a way that is incompatible with those purposes \cite{van2003handbook,solove2006taxonomy,dritsas2006knowledge}.  \textit{PoU} can be analyzed depending on the type of \textit{PoU}, if it is \textit{incompatible} a violation can be raised.

\end{description}

\item[\textbf{Anonymity,}] the identity of information owner should not be disclosed unless it is strictly required \cite{dritsas2006knowledge,solove2006taxonomy,ISO15408,Pfitzmann2010}, i.e., the primary/secondary identifiers of the data subject (e.g., name, social security number, address, etc.) should be removed if they are not strictly required and information still can be used for the same purpose after their removal.  Personal information can be \textit{anonymized} depending on some \textit{privacy mechanism}.

\item[\textbf{Unlinkability}]   means that it should not be possible to link personal information back to its owner. In other words, any identifiers that allow for such linkage should be removed \cite{ISO15408,mouratidis2007secure,kalloniatis2008addressing,Pfitzmann2010}. A \textit{privacy mechanism} can be used to remove any linkage between personal information and its owner.

\item[\textbf{Unobservability,}]  the identity of information owner should not be observed by others, especially third parties, while performing an activity (e.g., pursuing a goal) \cite{ISO15408,kalloniatis2008addressing,Pfitzmann2010}. Unlike \textit{Anonymity} and \textit{Unlinkability} that try to hide the identity of information owner, \textit{Unobservability} aims to hide some activities that are performed by the information owner \cite{Pfitzmann2010}. 

\textit{Unobservability} can be analyzed relying on the \textit{describes} relationship, which enables for detecting situations where personal information that describes an activity (goal) being pursued by a data subject is being collected by some other actor \cite{gharibsose17}.

\item[\textbf{Notice,}] information owner should be notified when its information is being collected \cite{van2003handbook,solove2006taxonomy,dritsas2006knowledge}.  \textit{Notice} is considered mainly to address situations where personal information related to a legitimate entity is being collected without her knowledge.  \textit{Notice} can be analyzed depending on the collect relationship and its corresponding permission. 

In case, personal information is being collected and there is no permission to collect, a notice violation will be raised. Providing a permission to collect means that the actor has been already notified and agrees his personal information to be collected.

\item[\textbf{Transparency,}] information owner should be able to know who is using his/her information and for what purposes \cite{van2003handbook,dritsas2006knowledge,kang2013security}. We rely on two principles to analyze transparency:

\begin{description}
\item[\textbf{Authentication}] a mechanism aims at verifying whether actors are who they claim they are. \textit{Authentication} can be analyzed by verifying whether 1- the actor is playing a role that enables to identify its main responsibilities; and 2- the actor is not playing any threat actor role. If both of these rules did not hold, a violation can be raised.

\item[\textbf{Authorization}] a mechanism aims at verifying whether actors can use information in accordance with their credentials \cite{dritsas2006knowledge}. \textit{Authorization} can be analyzed by verifying whether the actor has the required permissions to perform a task at hand.

\end{description}

\item[\textbf{Accountability,}] information owner should have a mechanism available to them to hold information users accountable for their actions concerning information  \cite{dritsas2006knowledge,kang2013security}. We rely on the \textit{non-repudiation} principle to analyze accountability:

\begin{description}
\item[\textbf{Non-repudiation,}] the delegatee cannot repudiate he/she accepted the delegation \cite{kang2013security}. \textit{Non-repudiation} can be analyzed relying on the adoption concept \cite{castelfranchi1998modelling},  if there exists a delegatee without an adopt relationship to the delegatum, a \textit{non-repudiation} violation can be raised.


\end{description}
\end{description}

\section{The implementation of COPri}

This section describes how we have implemented (codified) the COPri ontology depending on Prot\'eg\'e  software\footnote{\url{http://protege.stanford.edu/}}. In Prot\'eg\'e, ontology consists of Classes, Properties, and Individuals. \textit{Classes} are concrete representation of concepts, and they can be interpreted as sets that contain \textit{Individuals} (also known as instances of classes). In other words, \textit{classes} are used to specify conditions that must be satisfied by individuals to be a member of such classes, where all \textit{Classes} are subclasses of the class \textit{Thing}. While \textit{Properties} are binary relationships among \textit{Classes}/\textit{Individuals}.

We have implemented the conceptual model of COPri relying on classes and object properties in Prot\'eg\'e, and we have to modify and  create new classes and relationships during this process. We have also created new classes and subclasses to represent attributes of some classes, since classes in Prot\'eg\'e cannot have attributes. Moreover, for each class that has attributes with quantitative values, we have created a class (called a Value Partition pattern) to present each of these attributes, and several individuals to cover all quantitative values of each of the attributes.

For example, the class \textit{Sensitivity} that has \textit{Sensitivity level} attribute, which can have the following values\textit{Secret}, \textit{Sensitive}, \textit{Confidential} or \textit{Restricted} has been represented by a class named \textit{Sensitivity level} that has four defined individuals \textit{slSecret}, \textit{slSensitive}, \textit{slConfidential} or \textit{slRestricted}. Furthermore, we have defined the \textit{hasSensitive} property to link the \textit{PersonalInformation} class to the \textit{Sensitivity level} class.

\textit{Classes} may overlap and to ensure that an individual that belongs to one of the classes cannot be a member of any other class, such \textit{classes} must be made disjoint from one another. Thus, all \textit{primitive siblings} classes (e.g., PersonalInformation and PublicInformation) in our ontology have been made \textit{disjoint}. This helps the reasoner to check the logical consistency of the ontology. Moreover, we have used Probe Classes \cite{Horridge2011}, which are classes that are subclasses of two or more disjoint classes to test and ensure that the ontology does not include inconsistencies. 

Additionally, we have used a covering axiom to solve the open world assumption in OWL-based ontologies, where a covering axiom is a class that results from the union of the classes being covered. In other words, a covering axiom means that any member of the covered class must be a member of the classes being covered. For example, PersonalInformation and PublicInformation are the only subclasses of the Information class, and using a covering axiom here means that Information must be one of these two subclasses, i.e., Information is \textit{covered} by PersonalInformation and PublicInformation.

A restriction in Prot\'eg\'e can be used to describe a class of individuals based on the relationships that members of the class participate in. In other words, the class contains all of the individuals that satisfy the defined restriction. Restrictions can be categorized into existential and universal restrictions:

\begin{description}

\item[Existential restrictions] (also known as some restrictions (someValuesFrom) and denoted by $\exists$) describe classes of individuals that participate in \textit{at least} one relationship along a specified property to individuals that are members of a specified class \cite{Horridge2011}. 

\item[Universal restrictions] (also known as all restrictions (allValuesFrom) and denoted $\forall$) describe classes of individuals that for a given property \textit{only} have relationships along this property to individuals that are members of a specified class \cite{Horridge2011}. 

\end{description}

By relying on Existential restrictions, we could say that a class is a subclass of other class if \textit{some} property (relationship) holds. For example, PersonalInformation is a subclass of the Information class if \textit{some} \textit{related} property to the \textit{Actor} class exist.  This is called \textit{necessary conditions} and it means if something is a member of this class then it is \textit{necessary} to fulfill these conditions. However, with necessary conditions, we cannot say that, if something fulfills these conditions then it \textit{must} be a member of this class.

This problem can be solved by relying on \textit{sufficient conditions} that use universal restrictions, which means if something fulfills the defined conditions then it \textit{must} be a member of this class. In this context, \textit{sufficient conditions}  enable us to say that if something is a member of the class PersonalInformation it is necessary for it to be a kind of Information, and it is necessary for it to \textit{only} have a property of type \textit{related} to the \textit{Actor}  class. 

Using only \textit{sufficient conditions},   the class PersonalInformation may also contains individuals that are Information and do not participate in any property of type \textit{related} to the \textit{Actor} class because \textit{universal restrictions} do not specify the existence of a relationship. They merely state that if a relationship exists for the property then it must be to individuals that are members of a specific class.

This problem can be solved by using both of \textit{necessary and sufficient conditions}, which  enables to say, if something is a member of the class PersonalInformation, then it is necessary for it to be a kind of Information, and it is necessary for it to have a property of type \textit{related} to the \textit{Actor} class.  In other words, using both of the \textit{necessary and sufficient conditions} is sufficient to recognize all classes that must be a member of the class PersonalInformation.

On the other hand, properties are used to link individuals from a domain to individuals from a range. Thus, we have defined the domain and range for each object property in our ontology (shown in Table \ref{table:ObjProperties}), which can be used by a reasoner to make inferences and detect inconsistencies. For example, the domain of property (relationship) \textit{aims} is the class \textit{Actor} and its range is the class \textit{Goal}. 

Moreover, we used only one inverse property (e.g.,  related property between PersonalInformation and actor classes) in our ontology to minimize the number of properties and because most of such properties can be inferred. Finally, we have used cardinality restrictions to specify the number of relationships between classes depending on at \textit{least}, at \textit{most} or \textit{exactly} keywords. 

\begin{figure*}[!b]
\centering
\includegraphics[width= 0.99 \linewidth]{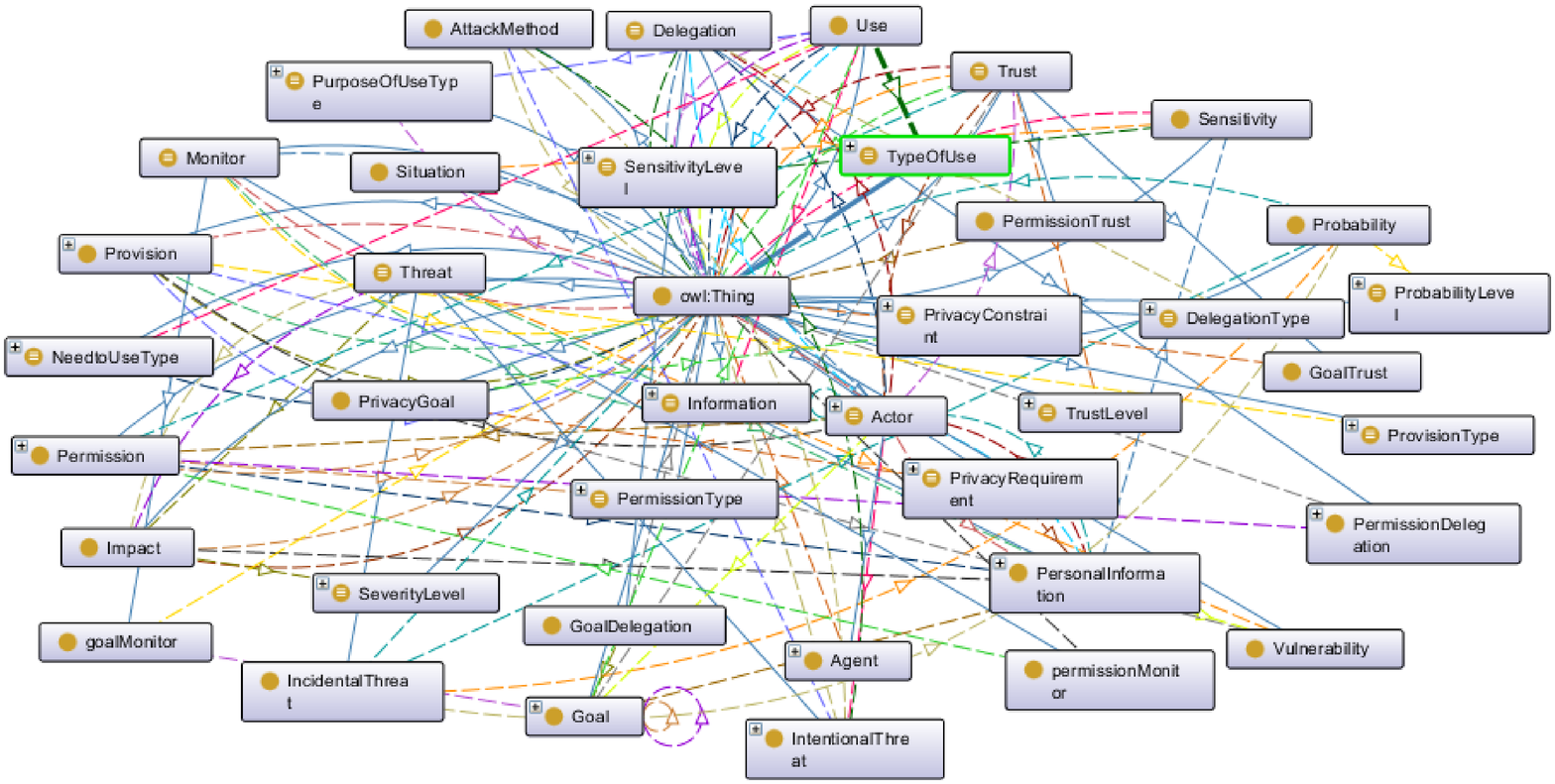}
\caption{A snapshot of the COPri ontology using OntoGraf plug-in}
\label{fig:OntoGraf}
\end{figure*}

\newpage
A snapshot of the COPri ontology is shown in Figure \ref{fig:OntoGraf}, and the COPri ontology is available in OWL formal at \url{https://goo.gl/AaqUxx}.

\begin{table}[!t]
\caption{Description of the domain and range of object properties}
\centering
\resizebox{\textwidth}{!}{
\begin{tabular}{p{2.7cm} | p{2.3cm} | p{2.3cm} || p{2.7cm} | p{2.3cm} | p{2.2cm}    }

\textbf{Object property} & \textbf{Domain}  &  \textbf{Range} & \textbf{Object property} & \textbf{Domain}  &  \textbf{Range} \\\hline 

                  adopts &   Actor          & Delegation      &   
							     aims  &   Actor          &    Goal          
									\\
          andDecomposed  &       Goal       &  Goal           &  
					   appliedTo     &   Pri.Mechanism     &    Per.Information  
					         \\
					concerning     &   Pri.Requirement   &  Per.Information   &  
						delegatee    &   Delegation     &    Actor      
						    	\\
            delegator    &        Actor     &   Delegation    &    
					describes      & Per.Information  &Goal       
						      \\
     determines          &  Situation       &  SensitivityLevel    &   
		 exploits            &    Threat        &  Vulnerability  
		              \\
       goalDelegatum     & goalDelegation   &    Goal         &  
		 goalTrustum         &     Trust        &    Goal        
		              \\
									
	 hasDelegationType     &   Delegation          &  DelegationType     &   
	
	            hasImpact  &      Threat &  Impact   \\
	
   hasNeedtoUseType      & Use                   &   NeedtoUseType    &  
		               hasPermission     &   Actor          &  Permission     \\
									
									
		hasPermissionType    &     Permission   &   PermissionType    &
	  hasProbability      & Inc.Threat        &   Probability 			\\

			hasProvisionType   & Provision        & ProvisionType   &
		
    hasPoUType  &    Use           &     PoUType              \\

    hasSensitivity    &  Per.Information &    SensitivityLevel &          
		hasSeverityLevel &   Impact   &   SeverityLevel 		\\

			hasTrustLevel  		 &    Trust     &   TrustLevel &     
      hasTypeOfUse         &      Use         &   TypeOfUse 	\\

			impactOver	 &    Impact        &  Per.Information        &
			
			includes			 &  Int.Threat &    AttackMethod     						\\

			intends   &    Actor        &   Int.Threat         &

      interpretedBy   &  Pri.Requirement  &   PrivacyGoal       \\   
			
			is\_a		 &     Role         &      Role          &
			isSubjectTo   & Per.Information  &  Vulnerability  					 \\   
			
			mitigates     	 &  PrivacyGoal     &   Vulnerability     &        
       monitor           &         Actor    &     Monitor     \\

			monitoree 				 &     Monitor     &    Actor &
      ofGoal      & goalMonitor       &  Goal             \\
				
			ofPermission		 & perm.Monitor &   permission        &      
       orDecomposed      &      Goal            &    Goal   \\
			
			over  			 &    Permission        &   Per.Information             &
         own       &     Actor         &    Per.Information             		\\
																	
						partOf  		 &    Information    &                Information				        &   				
     perm.Delegatum &    perm.Delegation              &     PermissionType    		 \\
		
		  perm.Trustum    &        Trust          &      Permission          &   
             plays       &    Agent              &      Role           \\

					provideTo			 &  Provision            &    Actor                 &   
          provideBy      &        Actor          &   Provision         		 \\

				provisionOf	  	 &   Provision           &   Information   &   
        realizedBy     &    PrivacyGoal  &   Pri.Constraint             \\								
				
				related	 & Per.Information    &       Actor         &   
        threaten            &      Threat             &   Per.Information   \\								
				
				trustee					 &    Trust              &        Actor   &    								
         trustor              &  Actor                &   Trust    \\								
				
				usedBy					 &    Goal              &      Use          &   
        usedOf        &      Use            &    Information            	 \\

\hline
\end{tabular}}
\label{table:ObjProperties}
\end{table}

\section{The validation of COPri}

\begin{figure*}[!t]
\centering
\includegraphics[width= 0.99 \linewidth]{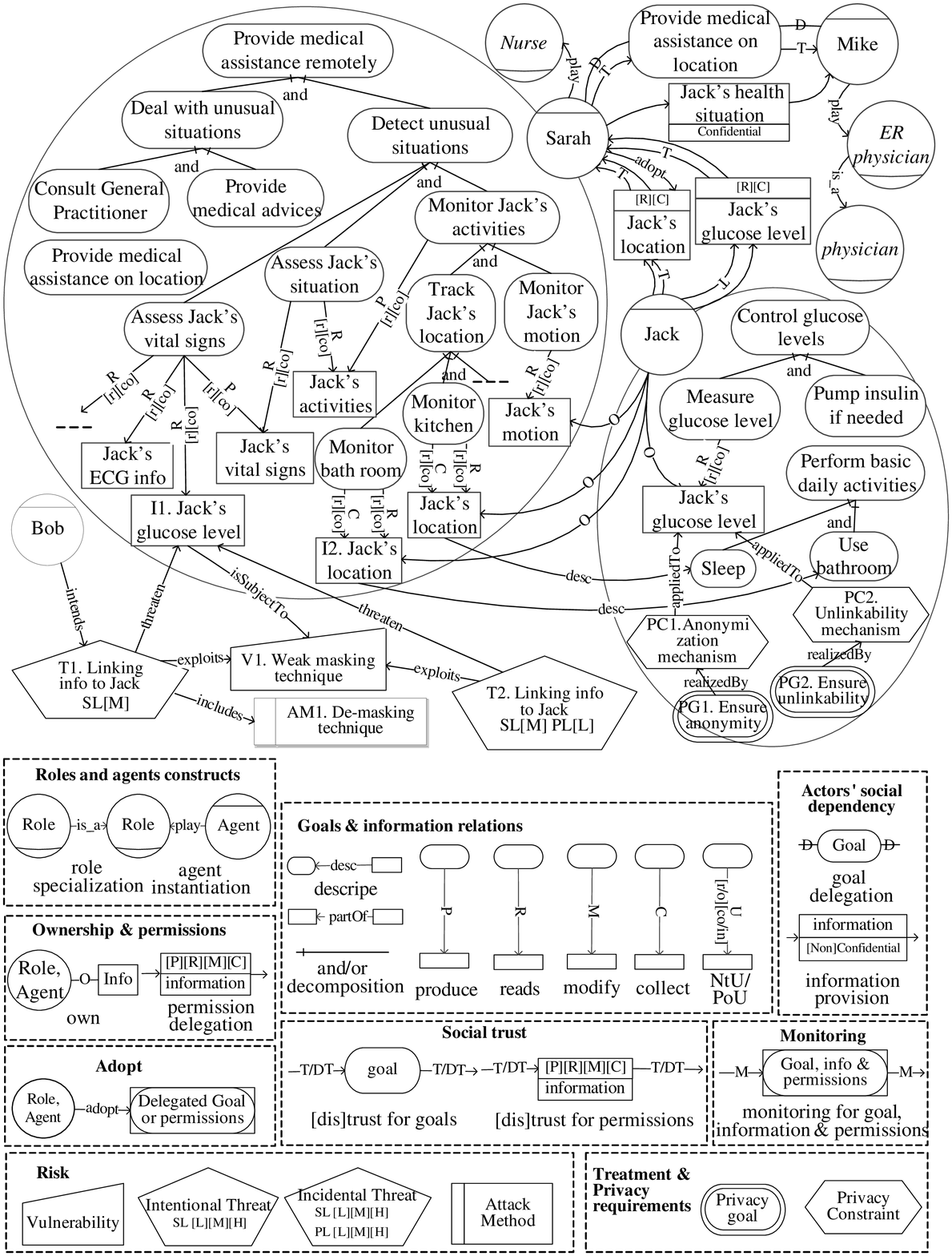}
\caption{A partial goal model concerning the AAL illustrating example}
\label{fig:Gore1}
\end{figure*}

In this section, we discuss how we validated our ontology depending on Competency Questions (CQs) that the ontology should be able to answer, i.e., to evaluate whether the ontology is able to capture detailed information about the targeted domain to fulfill the needs of its intended use \cite{fernandez1997methontology}. In other words, CQs represent a set of questions that the ontology must be capable of answering to be considered competent for tackling the problem it has been developed to solve \cite{Fox1993,uschold1996building,Dong2007}. In particular, we applied the COPri ontology to the AAL illustrating example, and then we validated COPri by formulating a set of CQs to query the ontology instances and check whether these queries are able to return reliable answers.  Figure \ref{fig:Gore1} shows a partial diagram of the AAL illustrating example represented in an extended goal model language\footnote{Note that this modeling language has not been developed yet, the diagram has been developed to assist the reader better understanding the usefulness of CQs}. 

The CQs are meant to assist and guide requirements engineers while dealing with privacy requirements in their social and organizational context by returning useful knowledge concerning the ontology. Moreover, some CQs have been developed to assist designers capturing (detecting and reporting) wrong/bad design decisions related to the four dimensions of our ontology, namely organizational, risk, treatment, and most importantly privacy requirements (e.g., confidentiality violation, notice violation, etc.). 

The formulation of the CQs was an iterative process and aimed at covering main wrong/bad design decisions that we call violations. Therefore, several CQs have been refined and extended before having the final set of CQs. Note that the concepts and relationships of the ontology have been refined and extended as well when we were formulating the CQs because some limitations and inadequacies in the ontology have been revealed. The set of CQs\footnote{Note that the main focus of the CQs is privacy requirements, not goal analysis} is shown in Table \ref{table:SPARQL}, where each CQ is represented both informally (natural language) and formally (SPARQL query). In what follows, we describe each of these four groups of CQs:  

\textbf{CQ1-3} are used to query organizational related aspects, where \textbf{CQ1} can be used to capture situations where a permission is delegated without a trust or trust compensation (e.g., monitoring). With the absence of trust and monitoring relationships, the delegator cannot guarantee that the delegatee will not misuse the delegated permission. For example, if there was no trust nor monitoring between Jack and Sarah concerning the delegation of read and/or collect permissions of Jack's location (shown in Figure \ref{fig:Gore1}), \textit{CQ1} will detect and report such situation. 

\textbf{CQ2} can be used to capture situations, where an actor monitors a delegation of permission although he/she trusts the delegatee, i.e., both trust and monitoring are used concerning the delegation of permission. In such situation, monitoring is not required and it is considered a bad design decision.  Concerning the previous example, if there is also a monitoring concerning the delegation of read/collect between Jack and Sarah, \textit{CQ2} will detect such situation and report that the monitoring relationship is not required.

\textbf{CQ3} can be used to return different sets of personal information based on their sensitivity levels (e.g., Secret, Sensitive, Confidential or Restricted).

\afterpage{
\begin{longtable}{p{1.1cm}  p{11cm}}

\caption{Competency Questions for validating the COPri ontology} 
\label{table:SPARQL}

\endfirsthead
\endhead

\hline

\multicolumn{2}{c}{\textbf{Organizational dimension}} \\  \hline

\textbf{CQ1.} &  Who are the delegators that delegate produce, read, modify, or collect permission, which is not accompanied by trust nor monitoring? \\   \cline{2-2}

     &   \texttt{SELECT} $?$actor1                     \\
	            &    \quad  \quad  \texttt{WHERE}              \{$?$actor1   copri:delegator   $?$delegate. \\
							&   \quad  \quad   \quad  \quad   \quad          $?$delegate copri:permissionDelegatum   copri:permProduce. \\
							&   \quad  \quad   \quad \quad     \quad  \quad   [:permRead   $|$  :permModify  $|$  :permCollect]                 \\
              &   \quad  \quad   \quad  \quad   \quad          $?$delegate   copri:delegatee   ?actor2.   \\

             	&   \quad   \quad    \texttt{FILTER NOT EXISTS} \{        $?$actor1   copri:trustor   ?trust.   \\
							&   \quad  \quad   \quad  \quad   \quad          $?$trust   copri:trustee   ?actor2.   \\
							&   \quad  \quad   \quad  \quad   \quad          $?$trust   copri:hasTrustLevel   copri:trust.   \\							
							&   \quad  \quad   \quad  \quad   \quad          $?$trust   copri:permissionTrustum    $?$perm.   \\	
							&   \quad  \quad   \quad  \quad   \quad          $?$perm   copri:hasPermissionType    copri:permProduce. \\
							&   \quad  \quad   \quad \quad     \quad  \quad   [:permRead   $|$  :permModify  $|$  :permCollect]\}               \\

							&   \quad   \quad    \texttt{FILTER NOT EXISTS} \{$?$actor1   copri:monitor   ?monitor.   \\
							&   \quad  \quad   \quad  \quad   \quad          $?$monitor   copri:monitoree    ?actor2.   \\
							&   \quad  \quad   \quad  \quad   \quad          $?$monitor   copri:ofPermission    $?$perm.\}\}  \\					\hline

\textbf{CQ2.}  &  Who are the delegators that delegate produce, read, modify, or collect permission accompanied by both trust and monitoring? \\     \cline{2-2}

     &   \texttt{SELECT} $?$actor1                  \\
	            &    \quad  \quad  \texttt{WHERE}              \{$?$actor1   copri:delegator   $?$delegate. \\
							&   \quad  \quad   \quad  \quad   \quad          $?$delegate copri:permissionDelegatum   copri:permProduce. \\
							&   \quad  \quad   \quad \quad     \quad  \quad   [:permRead   $|$  :permModify  $|$  :permCollect]                 \\
              &   \quad  \quad   \quad  \quad   \quad          $?$delegate   copri:delegatee   ?actor2.   \\
              &   \quad  \quad   \quad  \quad   \quad          $?$actor1   copri:trustor   ?trust.   \\
							&   \quad  \quad   \quad  \quad   \quad          $?$trust   copri:trustee   ?actor2.   \\
							&   \quad  \quad   \quad  \quad   \quad          $?$trust   copri:hasTrustLevel   copri:trust.   \\							
							&   \quad  \quad   \quad  \quad   \quad          $?$trust   copri:permissionTrustum    $?$perm.   \\	
							&   \quad  \quad   \quad  \quad   \quad          $?$perm   copri:hasPermissionType    copri:permProduce. \\
							&   \quad  \quad   \quad \quad     \quad  \quad   [:permRead   $|$  :permModify  $|$  :permCollect]               \\
							&   \quad  \quad   \quad  \quad   \quad          $?$actor1   copri:monitor   ?monitor.   \\
							&   \quad  \quad   \quad  \quad   \quad          $?$monitor   copri:monitoree    ?actor2.   \\
							&   \quad  \quad   \quad  \quad   \quad          $?$monitor   copri:ofPermission    $?$perm.\}    \\					\hline						
 
\textbf{CQ3.} &   Which are the personal information of sensitivity Restricted [Confidential, Sensitive or Secret]? \\     \cline{2-2}

              &   \texttt{SELECT} $?$PerInfo                   \\
	            &   \quad  \quad   \quad   \texttt{WHERE} \{$?$PerInfo copri:hasSensitivityLevel   copri:slRestricted \\
							&   \quad  \quad   \quad \quad  \quad   \quad  \quad  [:slConfidential   $|$  :slSensitive  $|$  :slSecret]\}      \\

\hline

\multicolumn{2}{c}{\textbf{Risk dimension}} \\  \hline

\textbf{CQ4.}  &  Which are the existing vulnerabilities and which personal information are subject to them? \\     \cline{2-2}

              &   \texttt{SELECT} $?$Vulnerability $?$PerInfo                  \\
	            &   \quad  \quad   \quad   \texttt{WHERE} \{$?$Vulnerability copri:isSubjectTo  $?$PerInfo\}      \\\hline

\textbf{CQ5.}   &   Which are the existing vulnerabilities and which are the threats that can exploit them? \\     \cline{2-2}

   &   \texttt{SELECT} $?$Threat $?$Vulnerability                  \\
	            &   \quad  \quad   \quad   \texttt{WHERE} \{$?$Threat copri:exploits  $?$Vulnerability\}      \\\hline

\textbf{CQ6.} &  Which are the existing vulnerabilities that are not mitigated by privacy goals?\\     \cline{2-2}

&   \texttt{SELECT} $?$Vulnerability                 \\
	            &   \quad  \quad   \quad   \texttt{WHERE} \{$?$Threat copri:exploits  $?$Vulnerability.        \\
							&   \quad  \quad   \quad     \texttt{FILTER NOT EXISTS}  \{?PriGoal    copri:mitigates $?$Vulnerability.\}       \\\hline

\textbf{CQ7.} &    Which are the existing threats and which are the personal information that are threatened by them? \\     \cline{2-2}

   &   \texttt{SELECT} $?$Threat $?$PerInfo                  \\
	            &   \quad  \quad   \quad   \texttt{WHERE} \{$?$Threat copri:threaten  $?$PerInfo\}      \\\hline

\textbf{CQ8.} &   Which are the existing threats that have an impact with severity level Low [Medium, High] over personal information? \\ \cline{2-2}

&   \texttt{SELECT} $?$Threat                 \\
	            &   \quad  \quad   \quad   \texttt{WHERE} \{$?$Threat copri:hasImpact  copri:SevLlowSeverity        \\
							&   \quad  \quad \quad   \quad \quad  \quad   \quad    [:SevLmediumSeverity $|$  :SevLhighSeverity]\}       \\\hline

\textbf{CQ9.}   &  Which  are the existing intentional threats and which are the personal information that are threatened by them? \\     \cline{2-2}

 &   \texttt{SELECT} $?$Threat $?$PerInfo                  \\
	            &   \quad  \quad   \quad   \texttt{WHERE} \{$?$Threat copri:threaten  $?$PerInfo.        \\
							&      \quad   \quad  \quad   \quad     \quad  \quad   \quad            $?$Threat copri:includes  $?$AttackMeth\}   \\\hline

\textbf{CQ10.}   &  Who are the threat actors and which are the intentional threats that they intend for? \\     \cline{2-2}

   &   \texttt{SELECT} $?$ThrActor $?$Threat                  \\
	            &   \quad  \quad   \quad   \texttt{WHERE} \{$?$ThrActor copri:intends  $?$Threat\}      \\\hline

\textbf{CQ11.}   &  Which are the existing attack methods and to which intentional threats they can be used for?\\     \cline{2-2}

   &   \texttt{SELECT}   $?$AttackMeth   $?$Threat              \\
	            &   \quad  \quad   \quad   \texttt{WHERE} \{$?$Threat copri:includes  $?$AttackMeth\}      \\\hline

\textbf{CQ12.}   &   Which are the existing incidental threats and which are the personal information that are threatened by them?\\     \cline{2-2}

 &   \texttt{SELECT} $?$Threat $?$PerInfo                  \\
	            &   \quad  \quad   \quad   \texttt{WHERE} \{$?$Threat copri:threaten  $?$PerInfo.        \\
							&      \quad   \quad  \quad   \quad     \quad  \quad   \quad            $?$Threat copri:hasProbability  $?$ProbLevel\}   \\\hline


\textbf{CQ13.} &   Which are the existing incidental threats of probability Low [Medium $|$ High]? \\ \cline{2-2}

&   \texttt{SELECT} $?$Threat                 \\
	            &   \quad  \quad   \quad   \texttt{WHERE} \{$?$Threat copri:hasProbability  copri:pllow         \\ 
							&   \quad  \quad  \quad   \quad \quad  \quad   \quad    [:plmedium $|$ :plhigh]\}      \\\hline

\multicolumn{2}{c}{\textbf{Treatment dimension}} \\  \hline

\textbf{CQ14.}  &  Which are the privacy goals that  are realized by privacy constraints? \\   \cline{2-2}

&   \texttt{SELECT} $?$PriGoal                 \\
	            &   \quad  \quad   \quad   \texttt{WHERE} \{$?$PriGoal copri:realizedBy  $?$PriCon\}       \\\hline

\textbf{CQ15.}  &  Which are the existing privacy mechanisms and which are the personal information that such mechanisms are applied to? \\   \cline{2-2}

              &   \texttt{SELECT} $?$PriMech $?$PerInfo                  \\
	            &   \quad  \quad   \quad   \texttt{WHERE} \{$?$PriMech copri:appliedTo  $?$PerInfo\}      \\\hline

\hline

\multicolumn{2}{c}{\textbf{Privacy dimension}} \\  \hline


\textbf{CQ16.}  & Which is the personal information that is read without read permission? \\     \cline{2-2}

&   \texttt{SELECT}  $?$PerInfo                \\
	            &   \quad  \quad      \texttt{WHERE}           \{$?$PerInfo     copri:related    $?$actor1.        \\
              &   \quad   \quad  \quad   \quad \quad  \quad    $?$PerInfo     copri:describes  $?$goal1.        \\
              &   \quad   \quad  \quad   \quad \quad  \quad    $?$use         copri:useOf      $?$PerInfo.        \\
              &   \quad   \quad  \quad   \quad \quad  \quad     $?$goal2       copri:useBy      $?$use.        \\
              &   \quad   \quad  \quad   \quad \quad  \quad     $?$use         copri:hasTypeOfUse copri:read.          \\
		 					&   \quad   \quad  \quad   \quad \quad  \quad    $?$actor       copri:aims      $?$goal2.        \\
							&   \quad   \quad    \texttt{FILTER NOT EXISTS}  \{$?$actor  copri:hasPermission  $?$perm.    \\
							&   \quad   \quad  \quad  \quad   \quad  \quad            $?$perm   copri:hasPermissionType copri:permRead\}\}   \\   		\hline

\textbf{CQ17.}  & Which is the personal information that is transferred relying on non-confidential provision? \\     \cline{2-2}

&   \texttt{SELECT} $?$PerInfo                 \\
	            &       \quad   \quad   \texttt{WHERE}     \{$?$PerInfo copri:related $?$Actor.         \\ 
						  &   \quad     \quad   \quad \quad  \quad   \quad 	$?$Prov copri:provisionOf $?$PerInfo.    \\ 
							&   \quad     \quad   \quad \quad  \quad   \quad   $?$Prov copri:hasProvisionType copri:nonConfidentialProv\}      \\\hline

\textbf{CQ18.}  & Which is the personal information that is used by a goal, where their usage (\textit{NtU}) are not strictly required (i.e., optional)? \\     \cline{2-2}

      &   \texttt{SELECT}  $?$PerInfo                  \\
	            &   \quad  \quad   \quad   \texttt{WHERE} \{$?$use copri:useOf  $?$PerInfo.  \\
						  &   \quad  \quad   \quad \quad  \quad   \quad    \quad       	$?$use   copri:hasNeedtoUseType copri:optional\}      \\\hline

\textbf{CQ19.}  & Which is the personal information that is used by goals, where their purpose of use (\textit{PoU}) is incompatible with the best interest of its owner? \\     \cline{2-2}

              &   \texttt{SELECT}  $?$PerInfo                  \\
	            &   \quad  \quad   \quad   \texttt{WHERE} \{$?$use copri:useOf  $?$PerInfo.  \\
						  &   \quad  \quad   \quad \quad  \quad   \quad    \quad       	$?$use    copri:hasPurposeOfUseType copri:incompatible\}      \\\hline

\textbf{CQ20.}  & Which is the personal information that is not anonymized? \\     \cline{2-2}

&   \texttt{SELECT} $?$PerInfo                \\
	            &   \quad  \quad   \quad   \texttt{WHERE} \{$?$PerInfo copri:related  $?$Actor.        \\
							&   \quad   \quad  \quad   \texttt{FILTER NOT EXISTS}  \{copri:PC1\_Anonymize  copri:appliedTo   \\
								&   \quad   \quad  \quad   \quad   \quad  \quad  \quad  \quad  \quad  \quad \quad  \quad  \quad     $?$PerInfo.\}\}   \\\hline

\textbf{CQ21.}  & Which is the personal information that can be linked back to their owners? \\     \cline{2-2}

							&   \texttt{SELECT} $?$PerInfo                \\
	            &   \quad  \quad   \quad   \texttt{WHERE} \{$?$PerInfo copri:related  $?$Actor.        \\
							&   \quad   \quad  \quad   \texttt{FILTER NOT EXISTS}  \{copri:PC2\_Unlinkability  copri:appliedTo.   \\
							&   \quad   \quad  \quad   \quad   \quad  \quad  \quad  \quad  \quad  \quad \quad  \quad  \quad $?$PerInfo\}\}   \\\hline

\textbf{CQ22.}  & Which is the personal information that describes a goal, and it is also being collected by some actor? \\     \cline{2-2}

				     	&   \texttt{SELECT} $?$PerInfo                \\
	            &   \quad  \quad   \quad   \texttt{WHERE}           \{$?$PerInfo     copri:related    $?$Actor.        \\
              &   \quad   \quad  \quad   \quad \quad  \quad   \quad $?$PerInfo     copri:describes  $?$goal1.        \\
              &   \quad   \quad  \quad   \quad \quad  \quad   \quad $?$use         copri:useOf      $?$PerInfo.        \\
              &   \quad   \quad  \quad   \quad \quad  \quad   \quad $?$goal2       copri:useBy      $?$use.        \\
              &   \quad   \quad  \quad   \quad \quad  \quad   \quad $?$use         copri:hasTypeOfUse copri:collect\}        \\\hline

\textbf{CQ23.}  & Who are the actors that are collecting personal information without collect permissions? (returns information too) ? \\     \cline{2-2}

&   \texttt{SELECT} $?$actor    $?$PerInfo            \\
	            &   \quad  \quad      \texttt{WHERE}           \{$?$PerInfo     copri:related    $?$actor1.        \\
              &   \quad   \quad  \quad   \quad \quad  \quad    $?$PerInfo     copri:describes  $?$goal1.        \\
              &   \quad   \quad  \quad   \quad \quad  \quad    $?$use         copri:useOf      $?$PerInfo.        \\
              &   \quad   \quad  \quad   \quad \quad  \quad     $?$goal2       copri:useBy      $?$use.        \\
              &   \quad   \quad  \quad   \quad \quad  \quad     $?$use         copri:hasTypeOfUse copri:collect.         \\
							&   \quad   \quad  \quad   \quad \quad  \quad    $?$actor       copri:aims      $?$goal2.        \\
							&   \quad   \quad        \texttt{FILTER NOT EXISTS}  \{$?$actor  copri:hasPermission  $?$perm.    \\
							&   \quad   \quad  \quad  \quad   \quad  \quad            $?$perm   copri:hasPermissionType copri:permCollect\}\}   \\\hline
							
\textbf{CQ24.}  &  Who are the actors that do not play any role or they are playing a threat actor role? \\     \cline{2-2}

              &   \texttt{SELECT} $?$actor                \\
              &   \quad  \quad      \texttt{WHERE}            \{\{$?$actor      rdf:type/rdfs:subClassOf*    copri:Agent.       \\
							&   \quad   \quad     \texttt{FILTER NOT EXISTS}  \{$?$actor  copri:plays  $?$role.\}\}    \\
							&   \quad   \quad     \texttt{UNION} \{$?$actor  copri:intends  $?$InThreat.\}\}    \\\hline

\textbf{CQ25.}  & Who are the actors that are using (producing, reading, modifying, or collecting) personal information without the required permission? (returns information too) ? \\     \cline{2-2}

&   \texttt{SELECT} $?$actor $?$PerInfo                \\
	            &   \quad  \quad      \texttt{WHERE}           \{$?$PerInfo     copri:related    $?$actor1.        \\
              &   \quad   \quad  \quad   \quad \quad  \quad    $?$PerInfo     copri:describes  $?$goal1.        \\
              &   \quad   \quad  \quad   \quad \quad  \quad    $?$use         copri:useOf      $?$PerInfo.        \\
              &   \quad   \quad  \quad   \quad \quad  \quad     $?$goal2       copri:useBy      $?$use.        \\
              &   \quad   \quad  \quad   \quad \quad  \quad     $?$use         copri:hasTypeOfUse copri:produce          \\
							&   \quad   \quad  \quad  \quad   \quad  \quad               [read $|$ modify $|$ collect]	\\
							&   \quad   \quad  \quad   \quad \quad  \quad    $?$actor       copri:aims      $?$goal2.        \\
							&   \quad   \quad      \texttt{FILTER NOT EXISTS}  \{$?$actor  copri:hasPermission  $?$perm.    \\
							&   \quad   \quad  \quad  \quad   \quad  \quad            $?$perm   copri:hasPermissionType copri:permProduce   \\
							&    \quad   \quad  \quad  \quad   \quad  \quad [:permRead $|$ :permModify $|$ :permCollect]\}\}   \\   		\hline

\textbf{CQ26.}  & Who are the delegatees that  have not adopted their delegatum? \\     \cline{2-2}

&   \texttt{SELECT} $?$actor                \\
	            &   \quad  \quad   \quad   \texttt{WHERE}\{\{$?$actor rdf:type/rdfs:subClassOf*  copri:Agent.        \\
							&   \quad   \quad  \quad      \texttt{FILTER NOT EXISTS}\{$?$actor  copri:plays $?$role.\}\}   \\
			        &   \quad  \quad   \quad   \texttt{UNION}\{$?$actor copri:intends  $?$InThreat\}\}    \\ \hline


\end{longtable}
}

\textbf{CQ4-13} are used to query risk related aspects, where \textbf{CQ4} can be used to return all vulnerabilities as well as the personal information that is subject to them. For example, \textit{CQ4} will return ``V1. Weak masking technique'' and ``I1. Jack's glucose level'' if applied to the AAL example (Figure \ref{fig:Gore1}).  \textbf{CQ5} can be used to return vulnerabilities as well as threats that can exploit such vulnerabilities. Concerning the AAL example, it will return ``V1.'' and both ``T1. Linking info to Jack SL[M]'' and ``T2. Linking info to Jack SL[M] PL[L]''.  \textbf{CQ6} can be used to return unmitigated vulnerabilities. In the AAL example, only one unmitigated vulnerability will be returned (e.g., ``V1.'').  \textbf{CQ7} can be used to return any threat that is threatening personal information as well as the threatened information. This \textit{CQ} will return both ``T1.'', ``T2.'' and ``I1.'' if applied to the AAL example.

\textbf{CQ8} can be used to return different sets of threats based on their severity levels (e.g., Low, Medium, or High). If \textit{CQ8} has been applied to return threats with medium severity levels, both ``T1.'' and ``T2.'' will be returned. Otherwise, it will return nothing, i.e., no threats with Low nor High severity level exist in the AAL example.   \textbf{CQ9} can be used to return intentional threats as well as the personal information threatened by them. For example, \textit{CQ9} will return ``T1.''  and ``I1.'' as  ``T1.''  is the only intentional threat in the example.  \textbf{CQ10} can be used to return threat actors and the intentional threats they intend for, applying \textit{CQ10} to our example, will return ``Bob'' and ``T1.'' since ``Bob'' is the only threat actor in our example.

\textbf{CQ11} can be used to return attack methods and the intentional threats they are used for. For instance, \textit{CQ11} will return ``AM1. De-masking technique''  and ``T1.'' if applied to the AAL example. \textbf{CQ12} can be used to return incidental threats and personal information that are threatened by them, and it will return ``T2.'' and ``I1.''  if applied to the AAL example.  \textbf{CQ13} can be used to return different sets of incidental threats based on their probability levels (e.g., Low, Medium, or High). If \textit{CQ13} has been applied to return incidental threats with a Low probability level, ``T2.'' will be returned. Otherwise, it will return nothing since no incidental threats with medium nor high probability exists in the example.

\textbf{CQ14-15} are used to query treatment-related aspects, where \textbf{CQ14} can be used to return privacy goals that have been realized by privacy constraints. Applying \textit{CQ14} to our example will return two privacy goals ``PG1. Ensure anonymity'' and ``PG2. Ensure unlinkability''.  \textbf{CQ15} can be used to return privacy mechanisms as well as the personal information such mechanisms are applied to. Applying \textit{CQ15} to the example will return both ``PC1. Anonymization mechanism'' and ``PC2. Unlinkability mechanism'' privacy mechanisms as well as ``I1.''

\textbf{CQ16-26} are used to query privacy requirements related violations, where the definitions of the seven main privacy requirements violations are presented in Table \ref{table:Oppo}.  In particular, \textbf{CQ16-19} are used for analyzing \textit{Confidentiality}, where \textbf{CQ16-17} are used for analyzing non-disclosure by detecting and reporting when personal information is read without the owner's permission (\textit{CQ16}), or it has been transferred relying on non-confidential transmission mean   (\textit{CQ17}).  For example, if Sarah did not have a read permission concerning ``Jack's glucose level'', \textit{CQ16} will detect and report such situation. While if ``Jack's health situation'' (personal information) is transferred to Mike relying on the non-confidential provision, \textit{CQ17} will detect and report such violation.

\begin{table}[!t]
\caption{Definitions of privacy requirements violations}
\centering
\resizebox{0.99\textwidth}{!}{
\begin{tabular} {p{2.3cm}  p{9cm} }

\hline 

\textbf{Pri. req.} &  \textbf{Privacy requirement violation definition} \\ 

\textit{Confidentiality} & \textit{Disclosure,} personal information is disclosed without the owner's consent (permission), personal information is used for tasks,  where it is not strictly required, and/or personal information is used for tasks that are incompatible with the specific, explicit, legitimate purposes, which has been permitted to be used for. \\

\hline

\textit{Anonymity} & \textit{Identifiability,}  the identity of information owner can be sufficiently identified, i.e., it can be disclosed.  \\

\hline

\textit{Unlinkability} & \textit{Linkability,}  the link between personal information and its owner can be sufficiently distinguished, i.e., it is possible to like personal information back to its owner.  \\

\hline

\textit{Unobservability} & \textit{Observability,}  the identity of information owner can be observed/detected by others while performing an activity.  \\

\hline

\textit{Notice} & \textit{Unnotified,}  personal information is being collected without notifying its owner.  \\

\hline

\textit{Transparent} & \textit{Untransparent,}   information owner is not able to know who is using his/her information and for what purposes.  \\

\hline

\textit{Accountable} & \textit{Unaccountable,}   information owner cannot hold information users accountable for their actions concerning its personal information.   \\

\hline
\end{tabular}}
\label{table:Oppo}
\end{table}

\textbf{CQ18} is used for analyzing Need to Know (NtK) principle by verifying whether personal information is strictly required by goals using them. For example, if the Need to Use (NtU) of the goal ``Assess Jack's situation'' concerning ``Jack's vital signs'' is optional, \textit{CQ18} will detect and report that such information is not strictly required by the goal. 

\textbf{CQ19} is used for analyzing the Purpose of Use (PoU) principle by verifying whether personal information is used for specific, explicit, legitimate purposes that have been permitted to be used for.  For instance, if the PoU of the goal ``Assess Jack's situation'' concerning ``Jack's vital signs'' is incompatible, \textit{CQ19} will detect and report that the PoU of such information is incompatible for specific, explicit, legitimate purposes that have been permitted to be used for.

\textbf{CQ20} is used for analyzing anonymity by verifying whether the identity of the information owner can be sufficiently identified. For example, if ``Jack's glucose level'' has not been anonymized relying on ``PC1. Anonymization mechanism'', \textit{CQ20} will detect and report that ``Jack's glucose level'' can be used to sufficiently identify the identity of Jack. 

\textbf{CQ21} is used for analyzing Unlinkability by verifying whether it is possible to link personal information back to its owner. For example, if ``PC2. Unlinkability mechanism'' was not applied to ``Jack's glucose level'', \textit{CQ21} will detect and report that it is possible to link ``Jack's glucose level'' to Jack.

\textbf{CQ22} is used for analyzing unobservability by verifying whether the identity of the information owner can be observed by others while performing some activity. Consider that Jack does not want his activities to be monitored while he is in the bathroom.  Then ``Jack's location'' should not be collected when he is in the bathroom,  since such information can be used to describe activities performed by Jack, which he does not want it to be observed by others. If ``Jack's location'' is collected, \textit{CQ22} will be able to detect and report that.

\textbf{CQ23} is used for analyzing notice by verifying whether personal information is being collected without notifying its owner. We consider that providing permission to collect implies that the actor has been already notified and agreed upon the collection of its personal information. In case, personal information is being collected and there is no permission to collect, \textit{CQ23} will detect and report such violation.

\textbf{CQ24-C25} are used for analyzing transparency, where \textbf{CQ24} analyze the authentication principle and \textbf{CQ25} analyze the authorization principle. In particular, \textbf{CQ24} verifies whether an actor can be authenticated by checking if it is playing at least one role that enables for identifying its main responsibilities\footnote{If an actor is not playing any role, it will be impossible to authenticate it}, and the actor is not playing any threat actor role. Accordingly, \textit{CQ24} will be able to detect and report whether such actor can be authenticated. Considering our example, Bob is playing a threat actor role, and it will be returned if \textit{CQ24} was applied. While \textbf{CQ25}  analyze authorization by verifying that actors are not using personal information without the required permissions. In case, Sarah was reading/collecting any of Jack' personal information without read/collect permission, \textit{CQ25} will be able to detect and report such violation.

Finally, \textbf{CQ26} is used for analyzing accountability relying on the non-repudiation principle by verifying that actors cannot repudiate that they accepted delegations, which can be done depending on the adoption concept, if there exists a delegatee without an adopt relationship to the delegatum, \textit{CQ26}  will detect and report such violation. Concerning our example, if Sarah did not adopt the read and collect permissions that have been delegated by Jack, \textit{CQ26} will detect and report such violations.

\section{Evaluation}

Evaluation aims to provide evidence that artifact achieves the purpose for which it has been developed \cite{hevner2004design,pries2008strategies,venable2012comprehensive,gregor2013positioning}.  We evaluate the COPri ontology against the common pitfalls in ontologies identified in \cite{Povedavillalon2010}\footnote{The catalog of the pitfalls can be found in \nameref{Appendix:A}}. These pitfalls can be classified by criteria under 1- \textit{Consistency} verifies whether the ontology includes or allows for any inconsistencies; 2- \textit{Completeness} verifies whether the domain of interest is appropriately covered; and 3- \textit{Conciseness} verifies whether the ontology includes irrelevant elements or redundant representations of some elements with respect to the domain to be covered.  The pitfalls classification by criteria is shown in Table \ref{table:pitfalls}, where we can also identify the four different methods we followed to evaluate the COPri ontology against each of the pitfalls:

\setlength\tabcolsep{1pt}
\begin{table*}[t]
  \centering

  \caption{Pitfalls classification by criteria and how they were evaluated}

		\hspace*{-58pt}\begin{tabular}{>{\raggedright}  c p{0.8cm} p{9cm} | p{0.3cm} | p{0.3cm} | p{0.3cm} | p{0.3cm} |}


\cline{3-7}  &  &   & \rotatebox{90}{Prot\'eg\'e} & \rotatebox{90}{OOPS!} & \rotatebox{90}{Experts}   & \rotatebox{90}{Researchers}  \\
 
\cline{1-7} \multicolumn{1}{ |c|  }{\multirow{9}{*}[-0.4ex]{\rotatebox{90}{\textbf{Consistency}}} }    &

                                           \textbf{P1.}  & Creating polysemous elements         &  \multicolumn{1}{ |c|  }{\textbf{-}}        &      \multicolumn{1}{ |c|  }{\textbf{-}}        &     \multicolumn{1}{ |c|  }{\textbf{\checkmark}}         &       \multicolumn{1}{ |c|  }{\textbf{-}}                                    \\

\cline{2-7} \multicolumn{1}{ |c|  }{}    &  \textbf{P5.} & Defining wrong inverse relationships  & \multicolumn{1}{ |c|  }{\textbf{-}}     &  \multicolumn{1}{ |c|  }{\checkmark}  &   \multicolumn{1}{ |c|  }{\textbf{-}}   &   \multicolumn{1}{ |c|  }{\textbf{-}}        \\

\cline{2-7} \multicolumn{1}{ |c|  }{}    &  \textbf{P6.} & Including cycles in the hierarchy   & \multicolumn{1}{ |c|  }{\textbf{\checkmark}}     & \multicolumn{1}{ |c|  }{\textbf{\checkmark}}  &   \multicolumn{1}{ |c|  }{\textbf{-}}   &   \multicolumn{1}{ |c|  }{\textbf{-}}        \\

\cline{2-7} \multicolumn{1}{ |c|  }{}    &  \textbf{P7.} & Merging different concepts in the same class  & \multicolumn{1}{ |c|  }{\textbf{-}}     &  \multicolumn{1}{ |c|  }{\textbf{\checkmark}}   &  \multicolumn{1}{ |c|  }{\textbf{\checkmark}}     &   \multicolumn{1}{ |c|  }{\textbf{-}}         \\

\cline{2-7} \multicolumn{1}{ |c|  }{}    &  \textbf{P14.} & Misusing ``allValuesFrom''   & \multicolumn{1}{ |c|  }{\textbf{\checkmark}}     &  \multicolumn{1}{ |c|  }{\textbf{-}}    &   \multicolumn{1}{ |c|  }{\textbf{-}}   &   \multicolumn{1}{ |c|  }{\textbf{-}}        \\

\cline{2-7} \multicolumn{1}{ |c|  }{}    &  \textbf{P15.} & Misusing ``not some'' and ``some not''   & \multicolumn{1}{ |c|  }{\textbf{\checkmark}}     &  \multicolumn{1}{ |c|  }{\textbf{-}}   &   \multicolumn{1}{ |c|  }{\textbf{-}}   &   \multicolumn{1}{ |c|  }{\textbf{-}}        \\

\cline{2-7} \multicolumn{1}{ |c|  }{}    &  \textbf{P18.} & Specifying too much the domain or the range   & \multicolumn{1}{ |c|  }{\textbf{\checkmark}}     &  \multicolumn{1}{ |c|  }{\textbf{-}}    &   \multicolumn{1}{ |c|  }{\textbf{-}}   &   \multicolumn{1}{ |c|  }{\textbf{-}}        \\

\cline{2-7} \multicolumn{1}{ |c|  }{}    &  \textbf{P19.} & Swapping intersection and union   & \multicolumn{1}{ |c|  }{\textbf{-}}     &  \multicolumn{1}{ |c|  }{\textbf{\checkmark}}   &   \multicolumn{1}{ |c|  }{\textbf{-}}   &   \multicolumn{1}{ |c|  }{\textbf{-}}        \\

\cline{2-7} \multicolumn{1}{ |c|  }{}    &  \textbf{P24.} & Using recursive definition   & \multicolumn{1}{ |c|  }{\textbf{-}}     & \multicolumn{1}{ |c|  }{\textbf{\checkmark}}   &  \multicolumn{1}{ |c|  }{\textbf{\checkmark}}   &   \multicolumn{1}{ |c|  }{\textbf{-}}        \\

\cline{1-7} \multicolumn{1}{ |c|  }{\multirow{7}{*}[-0.4ex]{\rotatebox{90}{\textbf{Completeness}}} }    &   

                                           \textbf{P4.} & Creating unconnected ontology elements        &  \multicolumn{1}{ |c|  }{\textbf{\checkmark}}        &     \multicolumn{1}{ |c|  }{\textbf{\checkmark}}        &      \multicolumn{1}{ |c|  }{\textbf{-}}          &       \multicolumn{1}{ |c|  }{\textbf{-}}                                    \\

\cline{2-7} \multicolumn{1}{ |c|  }{}    &  \textbf{P9.} & Missing basic information  & \multicolumn{1}{ |c|  }{\textbf{-}}     &   \multicolumn{1}{ |c|  }{\textbf{-}}    &   \multicolumn{1}{ |c|  }{\textbf{-}}  &  \multicolumn{1}{ |c|  }{\textbf{\checkmark}}       \\

\cline{2-7} \multicolumn{1}{ |c|  }{}    &  \textbf{P10.} & Missing disjointness    & \multicolumn{1}{ |c|  }{\textbf{\checkmark}}     &  \multicolumn{1}{ |c|  }{\textbf{\checkmark}}   &   \multicolumn{1}{ |c|  }{\textbf{-}}   &   \multicolumn{1}{ |c|  }{\textbf{-}}        \\

\cline{2-7} \multicolumn{1}{ |c|  }{}    &  \textbf{P11.} & Missing domain or range in properties  & \multicolumn{1}{ |c|  }{\textbf{\checkmark}}     &  \multicolumn{1}{ |c|  }{\textbf{\checkmark}}    &   \multicolumn{1}{ |c|  }{\textbf{-}}   &   \multicolumn{1}{ |c|  }{\textbf{-}}        \\

\cline{2-7} \multicolumn{1}{ |c|  }{}    &  \textbf{P12.} & Missing equivalent properties & \multicolumn{1}{ |c|  }{\textbf{-}}     &  \multicolumn{1}{ |c|  }{\textbf{\checkmark}}   &   \multicolumn{1}{ |c|  }{\textbf{-}}   &   \multicolumn{1}{ |c|  }{\textbf{-}}        \\

\cline{2-7} \multicolumn{1}{ |c|  }{}    &  \textbf{P13.} & Missing inverse relationships   & \multicolumn{1}{ |c|  }{\textbf{-}}     &  \multicolumn{1}{ |c|  }{\textbf{\checkmark}}    &   \multicolumn{1}{ |c|  }{\textbf{-}}   &   \multicolumn{1}{ |c|  }{\textbf{-}}        \\

\cline{2-7} \multicolumn{1}{ |c|  }{}    &  \textbf{P16.} & Misusing primitive and defined classes    & \multicolumn{1}{ |c|  }{\textbf{\checkmark}}     &\multicolumn{1}{ |c|  }{\textbf{-}}   &   \multicolumn{1}{ |c|  }{\textbf{-}}   &   \multicolumn{1}{ |c|  }{\textbf{-}}        \\

\cline{1-7} \multicolumn{1}{ |c|  }{\multirow{4}{*}[-0.4ex]{\rotatebox{90}{\textbf{Conciseness}}} }    &   

                                            \textbf{P2.} &  Creating synonyms as classes       &  \multicolumn{1}{ |c|  }{\textbf{-}}        &     \multicolumn{1}{ |c|  }{\textbf{\checkmark}}       &      \multicolumn{1}{ |c|  }{\textbf{\checkmark}}     &     \multicolumn{1}{ |c|  }{\textbf{-}}                                  \\

\cline{2-7} \multicolumn{1}{ |c|  }{}    &  \textbf{P3.} & Creating the relationship ``is'' instead of using ``subclassOf'', ``instanceOf'' or ``sameIndividual''   & \multicolumn{1}{ |c|  }{\textbf{-}}     &  \multicolumn{1}{ |c|  }{\textbf{\checkmark}}    &   \multicolumn{1}{ |c|  }{\textbf{-}}   &   \multicolumn{1}{ |c|  }{\textbf{-}}        \\

\cline{2-7} \multicolumn{1}{ |c|  }{}    &  \textbf{P17.} & Specializing too much a hierarchy    & \multicolumn{1}{ |c|  }{\textbf{\checkmark}}     &  \multicolumn{1}{ |c|  }{\textbf{-}}    &   \multicolumn{1}{ |c|  }{\textbf{\checkmark}}   &   \multicolumn{1}{ |c|  }{\textbf{-}}        \\

\cline{2-7} \multicolumn{1}{ |c|  }{}    &  \textbf{P21.} & Using a miscellaneous class  & \multicolumn{1}{ |c|  }{\textbf{\checkmark}}     &  \multicolumn{1}{ |c|  }{\textbf{\checkmark}}    &   \multicolumn{1}{ |c|  }{\textbf{\checkmark}}   &   \multicolumn{1}{ |c|  }{\textbf{-}}        \\

    \bottomrule
    \end{tabular}\hspace*{-58pt}%
  \label{table:pitfalls}%
\end{table*}%

\noindent \textbf{1- Prot\'eg\'e \& HermiT Reasoner\footnote{\url{http://www.hermit-reasoner.com/}}:} HermiT is the first publicly available OWL reasoner, and it can perform various automated checks such as consistency, satisfiability, etc. of OWL-based ontologies.  Both Prot\'eg\'e \& HermiT have been used to verify COPri against several pitfalls. In particular, HermiT is able to detect cycles in the hierarchy (\textit{P6.}), and \textit{P4.} has been verified depending on  OntoGraf plug-in that enables to visualize the ontology. Concerning \textit{P10.},  we have already made all \textit{primitive siblings} classes in our ontology \textit{disjoint}, i.e., no missing disjoint can be found in the ontology. We have manually checked whether the domain and range of all object properties have been defined (\textit{P11.}). Moreover, we verified \textit{P14.} depending on Probe Classes \cite{Horridge2011}, which can be used to test and ensure that the ontology does not include inconsistencies.  

COPri ontology cannot suffer from \textit{P15.} since we did not use complement operators to describe/define any of the classes.  All defined classes have been defined depending on both necessary and sufficient conditions, which makes the inferred hierarchy exactly the same as the asserted one.  The concepts of the ontology are general enough to avoid both  \textit{P17.} and \textit{P18.}, i.e., specializing too much hierarchy (\textit{P17.}), or specifying the domain and/or the range too much (\textit{P18.}). Additionally, when we need very specific concepts (e.g., attribute of a class), we have used individuals.  No miscellaneous class have been identified (\textit{P21.}), since the names of all classes and their sub-classes have been carefully chosen.

\noindent \textbf{2- Evaluation with OntOlogy Pitfall Scanner (OOPS!):} OOPS! is a web-based ontology evaluation tool\footnote{\url{http://oops.linkeddata.es/index.jsp}} for detecting common pitfalls in ontologies \cite{Group2009}.  The ontology can be uploaded to OOPS!, which return an evaluation report about the detected pitfalls, where each pitfall is described by its identifier, title, description, elements affected (e.g., classes, object properties,  or even the whole ontology) and an importance level.  There are three importance levels based on the impact that a pitfall may have on the ontology: 1- \textit{Critical:} it is crucial to correct the pitfall. Otherwise, the consistency, reasoning, applicability, etc. of the ontology could be affected; 2- \textit{Important:} it is not critical for functionality of the ontology, but it is important to be corrected; and 3- \textit{Minor:} it does not represent a problem, but correcting it makes the ontology better organized and user friendly.

\begin{figure*}[!t]
\centering
\includegraphics[width= 0.99 \linewidth]{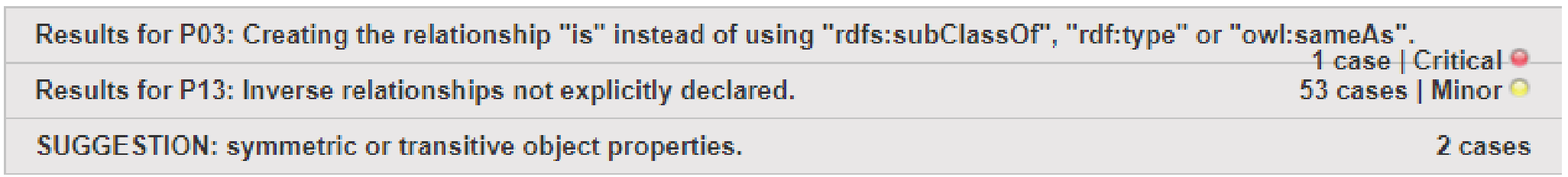}
\caption{OOPS! evaluation result}
\label{fig:OOPS1}
\end{figure*}

\begin{figure*}[!t]
\centering
\includegraphics[width= 0.99 \linewidth]{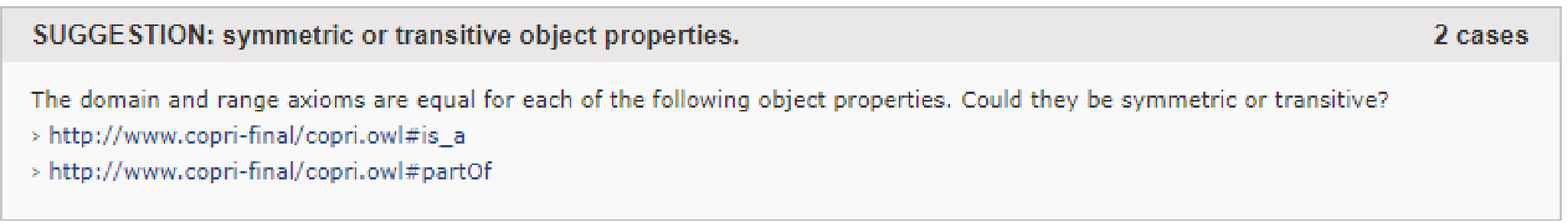}
\caption{Snapshot of the suggestions}
\label{fig:OOPSsug}
\end{figure*}

\noindent \textbf{Result of evaluation:}  the COPri ontology was uploaded to the OOPS! pitfall scanner, which returned an evaluation report that is shown in Figure \ref{fig:OOPS1}\footnote{Evaluation with OOPS! has been performed after evaluating the ontology with Prot\'eg\'e \& HermiT, i.e., several pitfalls have been already detected and corrected}.   In particular, two suggestions (Figure \ref{fig:OOPSsug}) have been returned, proposing that it might be better to characterize both of is\_a and partOf relationships as symmetric or transitive. We took these suggestions into account, characterizing both of these relationships as transitive.   53 minor pitfalls (\textit{P13}: inverse relationships not explicitly declared) have been identified. However, as mentioned earlier we used only one inverse property to minimize the number of properties/relationships in the ontology.

\begin{figure*}[!t]
\centering
\includegraphics[width= 0.99 \linewidth]{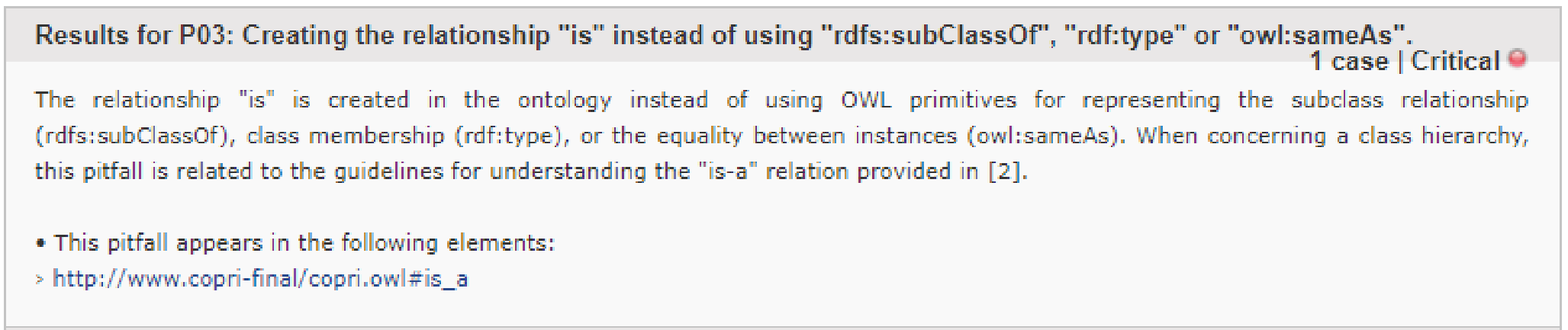}
\caption{Snapshot of the identified critical pitfall}
\label{fig:OOPSp03}
\end{figure*}

\begin{figure*}[!t]
\centering
\includegraphics[width= 0.99 \linewidth]{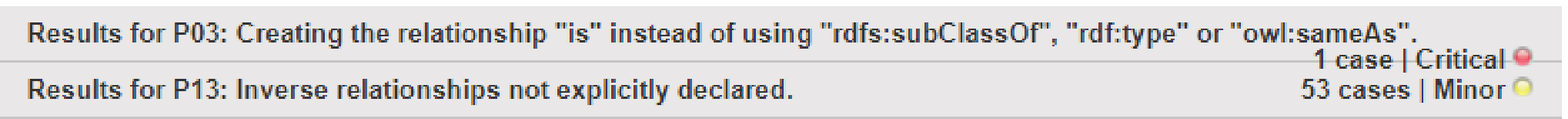}
\caption{OOPS! final evaluation result}
\label{fig:OOPS2}
\end{figure*}

Finally, only one critical pitfall has been identified shown in Figure \ref{fig:OOPSp03}, and looking closely at this pitfall,  we can see that the reseanor identify that we are using is\_a relationship instead of using OWL primitives for representing the subclass relationship (rdfs:subClassOf). However, is\_a relationship is used in most of Goal-based modeling languages, where we have adopted many of the concepts/relationships of the COPri ontology. Therefore, we chose not to replace it with the subClassOf relationship.  The result of the second test after addressing one of the suggestions is shown in Figure \ref{fig:OOPS2}.

\noindent \textbf{3- Lexical semantics experts:} two lexical semantics experts with main focus on Natural Language Processing (NLP) have been provided with the COPri ontology and they were asked to check whether the ontology suffers from any of the following pitfalls: \textit{P1.} Creating polysemous elements, \textit{P2.} Creating synonyms as classes, \textit{P7.} Merging different concepts in the same class, \textit{P17.} Specializing too much a hierarchy, \textit{P21.} Using a miscellaneous class, and \textit{P24.} Using recursive definition.

\noindent \textbf{Result of evaluation:}  several issues have been raised by the experts mainly concerning \textit{P2. Creating synonyms as classes}, and \textit{P24. Using recursive definition}. Most  of these issues has been properly addressed. The experts' feedback and how it was addressed can be found in \nameref{Appendix:B}.

\noindent \textbf{4- A survey with researchers:} the main purpose of this survey was evaluating the adequacy and completeness of the COPri ontology in terms of its concepts and relationships for dealing with privacy requirements in their social an organizational context (\textit{P9.}), i.e., whether the selected concepts and relationships are adequate to deal with privacy requirements or they need to be refined or extended.  The survey was closed, i.e., it was accessible through a special link that is provided to the invited participants only to avoid unintended participants. In what follows, we describe the survey participants, the survey template design, and the result of the survey:

\noindent \textbf{Survey participants:} in total 25 potential participants were contacted to complete the survey, and they were asked to forward the email to anyone who fits in the participating criteria (e.g., has good experience in privacy and/or security).  We have received 16 responses (64\% response rate).

\noindent \textbf{Survey template design:}  the survey template\footnote{The survey template can be found at \url{https://goo.gl/bro8nG}}, and it is composed of four main sections: \textit{S1. General information about the survey} includes a description of the purpose of the survey, privacy and confidentiality statement, and informed consent to be read and accepted (checked) by participants before providing any input. \textit{S2. Participant demographic} includes four questions related to the participant's name, occupation, type of experience (academic and/or industry), and years of experience with privacy and/or security. \textit{S3. Evaluating the COPri ontology} aims at collecting feedback from participates for evaluating the adequacy and completeness of the COPri ontology in terms of its main concepts and relationships categories and dimensions. \textit{S4. Final remarks}] aims at collecting suggestion and/or criticism concerning the COPri ontology.

\noindent \textbf{S2. Result of demographic questions.}   15 (93.8\%)  of the participants are researchers  (e.g., Professors, Post-docs, and PhD. candidates) and 1 (6.2\%)  is a student (e.g., MSc, bachelor). Concerning experience with privacy and/or security: 2 (12.5\%) of the participants have both academic and industrial experience, and 14 (87.5\%)  have pure academic experience. Moreover, 3 (18.8\%) have less than one year, 7 (43.8\%) have between one and four years, and 6 (37.5\%) have more than four years of experience. 

\noindent \textbf{S.3 Result of evaluation questions.}   This section is composed of 10 subsections, each of them is dedicated to collect feedback concerning the adequacy and completeness of a specific dimension/category of concepts and relationships.  In each of these subsections, we provide the definitions of the concepts and relationships of the targeted dimension/category as well as a diagram representing them. Followed by a mandatory question, asking the participant to grade the completeness of the presented concepts and relationships with respect to the system aspects they aim to capture on a scale from 1 (incomplete) to 5 (incomplete). 

In total, we have defined 10 questions each of them for a category or a dimension under evaluation. In particular, \textit{Q1-7} cover the seven main categories of concepts in the organizational dimension as follows: \textit{Q1} for the agentive entities category, \textit{Q2} for the intentional entities category, \textit{Q3} for the information entities category, \textit{Q4} for the goals \& information interrelationships category, \textit{Q5} for the information ownership \& permissions category, \textit{Q6} for the entities interactions category, and \textit{Q7}for the entities social trust category. Moreover, we defined \textit{Q8}for the risk dimension,  \textit{Q9} for the treatment dimension, and  \textit{Q10} for the privacy dimension.  The result of the evaluation for each of these sections is summarized in Table \ref{table:Qresult}.  The result tends to demonstrate that most of the targeted dimension/category of concepts and relationships are properly covering the aspects they aim to represent.

Additionally, we have added an optional question in each of 10 sections to evaluate the adequacy of the concepts and relationships by collecting suggestions to improve the category/dimension under evaluation.   Some feedback suggested to refine, include or exclude some of the concepts/relationships, we took some of these suggestions into account while developing the final ontology. 

\textbf{Result of the final remarks question:} most of the feedback was valuable, has raised important issues and ranged from complementing to criticizing.  For example, among the encouraging feedback we received \textit{``COPri covers a wide range of privacy-related concepts, with actor and goal oriented perspectives, which looks promising. We look forward to seeing it used to capture real-world privacy problem context''.} Another feedback and suggestion was \textit{``I think it is very precise and a very good work. Maybe some others concepts could be expressed somewhere''}. One of the comments we received was \textit{``How satisfaction of privacy requirements can be verified using it?''}. We also received criticisms such as the following one \textit {``I have no idea how good it is unless it is applied to many real cases.  I'm concerned that it is not grounded in reality.  It's also very complicated, which makes it hard to apply in industry''}. However, such criticism opens the way for future research directions.

\begin{table*}[!t]
\caption{The result of the evaluation}
\centering
\resizebox{\textwidth}{!}{
\begin{tabular}{p{3.2cm}  p{2cm}  p{2cm}p{2cm}p{2cm}p{2cm}}

              & \multicolumn{1}{c}{\textbf{Strongly}} &  \multicolumn{1}{c}{\textbf{Disagree}} & \multicolumn{1}{c}{\textbf{N. agree/}}   & \multicolumn{1}{c}{\textbf{Agree}}    & \multicolumn{1}{c}{\textbf{Strongly}} \\
              & \multicolumn{1}{c}{\textbf{disagree}} &  & \multicolumn{1}{c}{\textbf{n. disagree}} &            & \multicolumn{1}{c}{\textbf{agree}}    \\\hline																		
 \textit{Q1.} Agentive cat. & \multicolumn{1}{c}{0 (\%0)} &  \multicolumn{1}{c}{1 (\%6.3)} & \multicolumn{1}{c}{3 (\%18.8)}  & \multicolumn{1}{c}{6 (\%37.5)} & \multicolumn{1}{c}{6 (\%37.5)} \\

 \textit{Q2.} Intentional cat. & \multicolumn{1}{c}{0 (\%0)} &  \multicolumn{1}{c}{1 (\%6.3)} & \multicolumn{1}{c}{4 (\%25.0)}  & \multicolumn{1}{c}{7 (\%43.8)} & \multicolumn{1}{c}{4 (\%25.0)} \\

 \textit{Q3.} Informational cat. & \multicolumn{1}{c}{0 (\%0)} &  \multicolumn{1}{c}{2 (\%12.5)} & \multicolumn{1}{c}{4 (\%25.0)}  & \multicolumn{1}{c}{4 (\%25.0)} & \multicolumn{1}{c}{6 (\%37.5)} \\

 \textit{Q4.} Goals \& info cat. & \multicolumn{1}{c}{0 (\%0)} &  \multicolumn{1}{c}{2 (\%12.5)} & \multicolumn{1}{c}{2 (\%12.5)}  & \multicolumn{1}{c}{6 (\%37.5)} & \multicolumn{1}{c}{6 (\%37.5)} \\

 \textit{Q5.} Ownership cat. & \multicolumn{1}{c}{0 (\%0)} &  \multicolumn{1}{c}{1 (\%6.3)} & \multicolumn{1}{c}{1 (\%6.3)}  & \multicolumn{1}{c}{5 (\%31.3)} & \multicolumn{1}{c}{9 (\%56.3)} \\

 \textit{Q6.} Interactions cat. & \multicolumn{1}{c}{0 (\%0)} &  \multicolumn{1}{c}{1 (\%6.3)} & \multicolumn{1}{c}{1 (\%6.3)}  & \multicolumn{1}{c}{6 (\%37.5)} & \multicolumn{1}{c}{8 (\%50)} \\

 \textit{Q7.} Social Trust cat. & \multicolumn{1}{c}{0 (\%0)} &  \multicolumn{1}{c}{0 (\%0.0)} & \multicolumn{1}{c}{4 (\%25.0)}  & \multicolumn{1}{c}{7 (\%43.8)} & \multicolumn{1}{c}{5 (\%31.3)} \\

 \textit{Q8.} Risk  dim. & \multicolumn{1}{c}{0 (\%0)} &  \multicolumn{1}{c}{3 (\%18.8)} & \multicolumn{1}{c}{0 (\%00.0)}  & \multicolumn{1}{c}{8 (\%50)} & \multicolumn{1}{c}{5 (\%31.3)} \\

 \textit{Q9.} Treatment dim. & \multicolumn{1}{c}{0 (\%0)} &  \multicolumn{1}{c}{0 (\%0.00)} & \multicolumn{1}{c}{3 (\%18.8)}  & \multicolumn{1}{c}{7 (\%43.8)} & \multicolumn{1}{c}{6 (\%37.5)} \\

 \textit{Q10.} Privacy dim. & \multicolumn{1}{c}{0 (\%0)} &  \multicolumn{1}{c}{2 (\%12.5)} & \multicolumn{1}{c}{2 (\%12.5)}  & \multicolumn{1}{c}{5 (\%31.3)} & \multicolumn{1}{c}{7 (\%43.8)} \\\hline

\end{tabular}}
\label{table:Qresult}
\end{table*}

\section {Threats to validity}

After presenting and discussing the of our ontology, we list and discuss the threats to its validity in this section.  Following Runeson et al. \cite{runeson2009guidelines}, we classify the identified threats under  two types, internal and external:

\noindent  \textbf{1- Internal threats:} is concerned with factors that have not been considered in the study, and they could have influenced the investigated factors \cite{trochim2006research,runeson2009guidelines}. We have identified one threat:  \textit{Authors' background,}] the authors of this study have good experience in goal modeling (especially in \textit{i}* \cite{Yu:1995:MSR:922590} based languages). This may have influenced the selection and definitions of the concepts and relationships of the ontology. However, \textit{i}* languages have been developed with the aim to capture requirements in their social and organizational context, which is also a main objective of our ontology.

\noindent  \textbf{2-  External threats:} is concerned with to what extent the results of the study can be generalized \cite{runeson2009guidelines}.  We have identified two threats:  1. \textit{Validity of the survey result,} the number of participants can raise concerns about the validity of the result. However, most of them are experts with good experience in privacy, and some of them are high-profile researchers. 2. \textit{Extensive evaluation,} the ontology has been evaluated against the common pitfalls in ontologies with the help of some tools,  lexical semantics experts, and privacy researchers, yet it has not been applied in industry, which may reveal undetected errors and new ways to improve it. However, applying our ontology to real case studies from different domains is on our list for future work.

\section{Related work}

Several ontologies have been proposed for dealing with privacy and security. For example, Oltramari et al. \cite{Oltramari2018} propose PrivOnto, a semantic framework for the analyzing privacy policies, they also developed an interactive online tool that allows users to explore 23,000 annotated data practice instantiated in the PrivOnto knowledge base. Singhal and Wijesekera \cite{singhal2010ontologies} provide a security ontology that can be used to identify which threats endanger which assets and what countermeasures can be used. Moreover, Massacci et al. \cite{massacci2011extended} propose ontology for security requirements engineering that adopts concepts from Secure Tropos methodology \cite{mouratidis2007secure}, and several industrial case studies.   While Velasco et al. \cite{velasco2009modelling} introduce an ontology-based framework for representing and reusing security requirements based on risk analysis. Additionally, Kang and Liang \cite{kang2013security} developed security ontology for software development, which includes most common security concerns, and Dritsas et al. \cite{dritsas2006knowledge} developed an ontology for designing and developing a set of security patterns that can be used to deal with security requirements for e-health applications.  General privacy ontologies/taxonomies  (e.g., Anton and Earp \cite{anton2004requirements}, Solove et al. \cite{solove2006taxonomy}, and Wuyts et al. \cite{wuyts2009linking}) can serve as a general knowledge repository for a knowledge-based privacy goal refinement. 

On the other hand, several approaches for dealing with privacy requirements have been proposed in the literature. For instance, Spiekermann and Cranor \cite{spiekermann2009engineering} propose guidelines for building privacy-friendly systems and three-layer model of user privacy concerns and relate them to system operations in terms of data transfer, storage, and processing. In addition, they propose guidelines for building privacy-friendly systems. Moreover, Deng et al. \cite{deng2011privacy}  provide a methodology for modeling privacy-specific threats for software systems along with a catalog of privacy-specific threat tree patterns, which can be used to address threats identified during the analysis. Radics et al. \cite{radics2013preprocess} introduce the PREprocess, a framework for privacy requirements engineering, which has been designed to guide a privacy analyst during the collection and elicitation of privacy requirements through the identification of privacy-related patterns. 

Moreover, Labda et al. \cite{labda2014modeling} propose a privacy-aware Business Processes (BP) framework for modeling, reasoning and enforcing privacy constraints. The framework offers five concepts that can be used for analyzing privacy-related aspects such as access control,  separation of Tasks (SoT), Binding of Tasks (BoT), user consent, Necessity to know (NtK), etc.  Hong et al. \cite{hong2004privacy} propose a privacy risk model specifically for ubiquitous computing, which captures privacy concerns at high abstraction level, and then refining them into concrete specific solutions.  Gharib et al. \cite{gharib2016privacy} propose a holistic approach for analyzing privacy requirements that aim at assisting software engineers in designing privacy-aware systems by providing guidance and support while dealing with privacy requirements. Finally, Kalloniatis et al. \cite{kalloniatis2008addressing} introduce PriS that is a security requirements engineering method, which supports eight types of \textit{privacy goals} corresponding to the eight privacy concerns they identify in their work, namely: authentication, authorization, identification, data protection, anonymity, pseudonymity, unlinkability, and unobservability.

\section{Conclusions and Future Work}

We introduce COPri, a Core Ontology for Privacy requirements engineering that adopts and extends our previous work, where we proposed a privacy ontology that has been mined through a systematic literature review. In this paper, we extend and refine the concepts and relationships proposed in \cite{Gharib2017ER}, we have also implemented the ontology depending on Prot\'eg\'e, and applied it to illustrating example concerning  Ambient-Assisted Living systems. Then, we have validated the ontology by querying the ontology instance (AAL example) depending on Competency Questions (CQs). This allows evaluating whether the ontology is able to capture detailed knowledge about the targeted domain to fulfill the needs of its intended use. Finally, we evaluated the ontology against common pitfalls in ontologies with the help of some software tools,  lexical semantics experts, and privacy and security researchers. 

The main aim of developing COPri is assisting software engineers while designing privacy-aware systems by providing a generic and expressive set of key privacy concepts and relationships, which enable for capturing privacy requirements in their social and organizational context. This work is our second step towards proposing a well-defined privacy ontology, which when completed would constitute a great step forward in improving the quality of privacy-aware systems. However, much work is still to be done.

For future work, we plan to better validate our ontology by deploying it to capture privacy requirements for real case studies from different domains. Moreover, we will refine and analyze several privacy-related concepts. For example, we plan to better analyze how the sensitivity level of personal information can be determined based on the situation, and how sensitivity levels can be used to facilitate the identification of related privacy requirements. Moreover, we will refine the analysis of the \textit{Need to Use (NtU)} property, trying to better characterize the relationship between a goal and personal information. Additionally, a special attention will be given for refining the analysis of the \textit{Purpose of Use (PoU)} property, as \textit{Compatible}/\textit{Compatible} are two abstract to characterize such important property, and we will investigate how the \textit{PoU} can be determined automatically based on the characteristics of goal.

On the other hand, we are planning to develop a goal-oriented framework based on our ontology. This framework will be used for modeling and analyzing privacy requirements in their social and organizational context. Moreover, it will provide mechanisms for deriving the final privacy specifications in terms of privacy policies. This requires achieving two goals, defining privacy policy specification language and a set of rules for the automated derivation of privacy policy specifications from the requirements model. Finally, we aim to promote the adoption of our ontology by providing illustration and documentation as it is available only as a raw OWL file. This may encourage other researchers to adopt, use and extend or provide us with useful feedback.



\newpage


\section*{Appendix A} 
\label{Appendix:A}

\textbf{Catalog of Common Pitfalls} In what follows, we present the catalog of 24 pitfalls identified in \cite{Povedavillalon2010}:

\begin{description}

\item[\textbf{P1.}] \textbf{Creating polysemous elements:} an ontology element whose name has different meanings is included in the ontology to represent more than one conceptual idea. 

\item[\textbf{P2.}] \textbf{Creating synonyms as classes:} several classes whose identifiers are synonyms are created and defined as equivalent. For example, we could define ``Car'', ``Motorcar'' and ``Automobile'' as equivalent classes. This pitfall is related to the guidelines presented in \cite{Guarino2000}.

\item[\textbf{P3.}] \textbf{Creating the relationship ``is'' instead of using ``subclassOf'', ``instanceOf'' or ``sameIndividual'':}  the ``is'' relationship is created in the ontology instead of using OWL primitives for representing the subclass relationship (``subclassOf''), the membership to a class (``instanceOf''), or the equality between instances (``sameAs''). This pitfall is also related to the guidelines for understanding the ``is-a'' relation provided in \cite{Guarino2000}. 

\item[\textbf{P4.}] \textbf{Creating unconnected ontology elements:} ontology elements (classes, relationships or attributes) are created with no relation to the rest of the ontology. An example of this type of pitfall is to create the relationship ``memberOfTeam'' and to miss the class representing teams; thus, the relationship created is isolated in the ontology.

\item[\textbf{P5.}] \textbf{Defining wrong inverse relationships:} two relationships are defined as inverse relations when actually they are not. For example, something is sold or something is bought; in this case, the relationships ``isSoldIn'' and ``isBoughtIn'' are not inverse.

\item[\textbf{P6.}] \textbf{Including cycles in the hierarchy \cite{Gangemi2006,Guarino2000}:} a cycle between two classes in the hierarchy is included in the ontology, although it is not intended to have such classes as equivalent. That is, some class A has a subclass B and at the same time B is a subclass of A. An example of this type of pitfall is represented by the class ``Professor'' as a subclass of ``Person'', and the class ``Person'' as a subclass of ``Professor''.

\item[\textbf{P7.}] \textbf{Merging different concepts in the same class:} a class is created whose identifier is referring to two or more different concepts. An example of this type of pitfall is to create the class ``StyleAndPeriod'', or ``ProductOrService''.

\item[\textbf{P8.}] \textbf{Missing annotations:} ontology terms lack annotations properties. This kind of properties improves the ontology understanding and usability from a user point of view.

\item[\textbf{P9.}] \textbf{Missing basic information:} needed information is not included in the ontology. Sometimes this pitfall is related to the requirements in the Ontology Requirements Specification Document (ORSD)  \cite{Suarez-Figueroa2010} that are not covered by the ontology. Other times it is related to knowledge that could be added to the ontology in order to make it more complete. 

\item[\textbf{P10.}] \textbf{Missing disjointness \cite{Guarino2000,Gangemi2006,Presutti2008}:} the ontology lacks disjoint axioms between classes or between properties that should be defined as disjoint. For example, we can create the classes ``Odd'' and ``Even'' (or the classes ``Prime'' and ``Composite'') without being disjoint; such representation is not correct based on the definition of these types of numbers.

\item[\textbf{P11.}] \textbf{Missing domain or range in properties:} relationships without domain or range (or none of them) are included in the ontology. There are situations in which the relation is very general and the range should be the most general concept ``Thing''. This pitfall is related to the common error when defining ranges and domains described in \cite{Presutti2008}.

\item[\textbf{P12.}] \textbf{Missing equivalent properties:} when an ontology is imported into another, classes that are duplicated in both ontologies are normally defined as equivalent classes. However, the ontology developer misses the definition of equivalent properties in those cases of duplicated relationships and attributes. For example, the classes ``CITY'' and ``City'' in two different ontologies are defined as equivalent classes; however, relationships ``hasMember'' and ``has-Member'' in two different ontologies are not defined as equivalent relations.

\item[\textbf{P13.}] \textbf{Missing inverse relationships:} there are two relationships in the ontology that should be defined as inverse relations. For example, the case in which the ontology developer omits the inverse definition between the relations ``hasLanguageCode'' and ``isCodeOf'', or between ``hasReferee'' and ``isRefereeOf''.

\item[\textbf{P14.}] \textbf{Misusing ``allValuesFrom'' \cite{Presutti2008}:} this pitfall can appear in two different ways. In the first, the anomaly is to use the universal restriction (``allValuesFrom'') as the default qualifier instead of using the existential restriction (``someValuesFrom''). This means that the developer thinks that ``allValuesFrom'' implies ``someValuesFrom''. In the second, the mistake is to include ``allValuesFrom'' to close off the possibility of further additions for a given property. 

\item[\textbf{P15.}] \textbf{Misusing ``not some'' and ``some not''  \cite{Presutti2008}}: to mistake the representation of ``some not'' for ``not some'', or the other way round. An example of this type of pitfall is to define a vegetarian pizza as any pizza which both has some topping that is not meat and also has some topping that is not fish. This example is explained in more detail in \cite{Presutti2008}.

\item[\textbf{P16.}] \textbf{Misusing primitive and defined classes \cite{Presutti2008}:} to fail to make the definition ``complete'' rather than ``partial'' (or ``necessary and sufficient'' rather than just ``necessary''). It is critical to understand that, in general, nothing will be inferred to be subsumed under a primitive class by the classifier. This pitfall implies that the developer does not understand the open world assumption. A more detailed explanation and examples can be found in \cite{Presutti2008}. 

\item[\textbf{P17.}] \textbf{Specializing too much a hierarchy:} the hierarchy in the ontology is specialized in such a way that the final leaves cannot have instances, because they are actually instances and should have been created in this way instead of being created as classes. Authors in  \cite{Guarino2000} provide guidelines for distinguishing between a class and an instance when modeling hierarchies. An example of this type of pitfall is to create the class ``RatingOfRestaurants'' and the classes ``1fork'', ``2forks'', and so on, as subclasses instead of as instances. Another example is to create the classes ``Madrid'', ``Barcelona'', ``Sevilla'', and so on as subclasses of ``Place''. This pitfall could be also named ``Individuals'' are not Classes.

\item[\textbf{P18.}] \textbf{Specifying too much the domain or the range  \cite{Guarino2000,Presutti2008}:} not to find a domain or a range that is general enough. An example of this type of pitfall is to restrict the domain of the relationship ``isOfficialLanguage'' to the class ``City'', instead of allowing also the class ``Country'' to have an official language or a more general concept such as ``GeopoliticalObject''.

\item[\textbf{P19.}] \textbf{ Swapping intersection and union:} the ranges and/or domains of the properties (relationships and attributes) are defined by intersecting several classes in cases in which the ranges and/or domains should be the union of such classes. An example of this type of pitfall is to create the relationship ``takesPlaceIn'' with domain ``OlympicGames'' and with the range the intersection of the classes ``City'' and ``Nation''. This pitfall is related to the common error appear in \cite{Presutti2008,Guarino2000}.

\item[\textbf{P20.}] \textbf{Swapping Label and Comment:} the contents of the Label and Comment annotation properties are swapped. An example of this type of pitfall is to include in the Label annotation of the class ``Crossroads'' the following sentence ``the place of intersection of two or more roads''; and to include in the Comment annotation the word ``Crossroads''.

\item[\textbf{P21.}] \textbf{Using a miscellaneous class:} to create in a hierarchy a class that contains the instances that do not belong to the sibling classes instead of classifying such instances as instances of the class in the upper level of the hierarchy. This class is normally named ``Other'' or ``Miscellaneous''. An example of this type of pitfall is to create the class ``HydrographicalResource'', and the subclasses ``Stream'', ``Waterfall'', etc., and also the subclass ``OtherRiverElement''. 

\item[\textbf{P22.}] \textbf{Using different naming criteria in the ontology:}  no naming convention is used in the identifiers of the ontology elements. Some notions about naming conventions are provided in \cite{Guarino2000}. For example, we can name a class by starting with upper case, e.g. ``Ingredient'', and its subclasses by starting with lower case, e.g. ``animalorigin'', ``drink'', etc.

\item[\textbf{P23.}] \textbf{Using incorrectly ontology elements:}  an ontology element (class, relationship or attribute) is used to model a part of the ontology that should be modeled with a different element. A particular case of this pitfall regarding the misuse of classes and property values is addressed in  \cite{Guarino2000}. An example of this type of pitfall is to create the relationship ``isEcological'' between an instance of ``Car'' and the instance ``Yes'' or ``No'', instead of creating the attribute ``isEcological'' whose range is Boolean.

\item[\textbf{P24.}] \textbf{Using recursive definition:}  an ontology element is used in its own definition. For example, it is used to create the relationship ``hasFork'' and to establish as its range the following the set of restaurants that have at least one value for the relationship ``hasFork''.

\end{description}

\newpage

\section*{Appendix B} 
\label{Appendix:B}

The list of the pitfalls identified by experts are shown in Table \ref{table:Experts}. Each pitfall is described with its identifier (e.g., P1., P2., P7., P17., P21. or P24.), affected element(s) (e.g., a class or a relationship), a description of the pitfall, followed by how we addressed it.

\begin{longtable}{p{0.8cm} p{6cm} | p{6cm} }

\caption{Expert's comments and how they were addressed} 
\label{table:Experts}

\endfirsthead
\endhead

\hline

\textbf{ID} &  \textbf{Affected element} & \textbf{Description} \\  \hline

P2.  & ``A goal is a state of affairs that an actor  intends (aims) to achieve''.  &  The term ``intends'' might be confused with the term ``intends'' in the definition of the threat actor \\ \cline{2-3}

& \multicolumn{2}{l}{We have refined the definition of the goal to address this comment as follows: } \\

& \multicolumn{2}{l}{``A goal is a state of affairs that an actor aims to achieve.''} \\  \hline

P24.  & ``Information represents any informational entity without intentionality.'' & Information is used in its own definition \\  \cline{2-3}

& \multicolumn{2}{l}{We have refined the definition of Information as follows:} \\

& \multicolumn{2}{l}{``Information represents a statement provided or learned about something} \\
& \multicolumn{2}{l}{or someone.''} \\  \hline

P2.  & ``We adopt four different sensitivity levels that range from 1 to 4, where 4 is the most sensitive.''  & There is no need to include numerical levels, such information is already presented as natural language describing sensitivity levels. \\ \cline{2-3}

& \multicolumn{2}{l}{We have addressed this comment as follows: } \\

& \multicolumn{2}{l}{``We adopt four different sensitivity levels ordered as  \textit{(R)estricted}, \textit{(C)onfidential}, } \\

& \multicolumn{2}{l}{\textit{(S)ensitive}, and \textit{Secre(T)}, where \textit{Secre(T)} is the most sensitive.''} \\  \hline

P2.  & ``Accordingly, we adopt four corresponding categories (we represent as classes) of personal information, namely \textit{Restricted}, \textit{Confidential}, \textit{Sensitive}, and \textit{Secret}.'' & Adding \textit{Restricted}, \textit{Confidential}, \textit{Sensitive}, and \textit{Secret} subclasses of personal information is not required, such information is already captured by the sensitivity levels of personal information. \\   \cline{2-3}

& \multicolumn{2}{l}{``We have addressed this comment by removing the four subclasses of personal  } \\

& \multicolumn{2}{l}{information (e.g., \textit{Restricted}, \textit{Confidential}, \textit{Sensitive}, and \textit{Secret}).''} \\  \hline

P24.  & ``\textbf{Compliant} indicates that the purpose for which information is used is compliant with the rules that guarantee the best interest of its owner;'' (same for \textbf{Incompliant}) & Compliant/Incompliant are used in their own definitions\\  \cline{2-3}

& \multicolumn{2}{l}{We have performed the following modifications:} \\

& \multicolumn{2}{l}{``\textbf{Compatible} indicates that the purpose for which information is used is compliant} \\ 

& \multicolumn{2}{l}{with the rules that guarantee the best interest of its owner''} \\ 

& \multicolumn{2}{l}{``\textbf{Incompatible} indicates that the purpose for which information is used is not } \\ 

& \multicolumn{2}{l}{compliant with the rules that guarantee the best interest of its owner''} \\  \hline

P24.  &``\textbf{Describes} is a relationship between information and goal, where information describes the goal while it is pursued by some actor.'' & Describes is used in its own definition \\  \cline{2-3}

& \multicolumn{2}{l}{We have refined the definition of Describes to address this comment as follows: } \\

& \multicolumn{2}{l}{``\textbf{Describes} is a relationship between information and a goal, where information } \\

& \multicolumn{2}{l}{characterizes the goal while it is being pursued by some actor''} \\  \hline

P24.  &``\textbf{Information provision} captures the provision of (provisionOf) information ..'' & provision is used in its own definition \\  \cline{2-3}

& \multicolumn{2}{l}{We have addressed this comment as follows: } \\

& \multicolumn{2}{l}{``\textbf{Information provision} captures the transmission of information ..'' } \\\hline

P21.  & Information, personal and public information & Dividing information into public information and personal information indicates that a personal information cannot be public, which is not correct (the properties public and personal are not disjoint). I think that the sub classes should be public information and private information \\ \cline{2-3}

& \multicolumn{2}{l}{We did not made any changes to address this comment since we believe the } \\

& \multicolumn{2}{l}{concepts we adopt (e.g., public and personal)  are fine, and they are highly} \\

& \multicolumn{2}{l}{adopted and used by privacy researchers.} \\\hline

\end{longtable}

\end{document}